\documentclass[12pt]{article}
\input epsf.sty
\def\ZZZ{{\hbox{ Z\kern-1.6mm Z}}}
\newcommand{\rrho}{r}
\newcommand{\bA}{{\bf A}}
\newcommand{\bG}{{\bf G}}
\newcommand{\bF}{{\bf F}}

\newcommand{\eps}{\epsilon}

\newcommand{\AAA}{{\bf a}}

\newcommand{\ggg}{{\bf g}}
\newcommand{\fff}{{\bf f}}
\newcommand{\ccc}{{\bf c}}

\newcommand{\bC}{{\bf C}}
\newcommand{\OO}{{\cal O}}

\newcommand{\LL}{{\cal L}}

\newcommand{\wt}{\widetilde}
\newcommand{\wh}{\widehat}

\newcommand{\bd}{\bar{\rm D}}

\newcommand{\TT}{{\cal T}}

\newcommand{\be}{\begin{equation}}
\newcommand{\ee}{\end{equation}}
\newcommand{\ben}{\begin{eqnarray}\displaystyle}
\newcommand{\een}{\end{eqnarray}}
\newcommand{\refb}[1]{(\ref{#1})}
\newcommand{\p}{\partial}
\newcommand{\sectiono}[1]{\section{#1}\setcounter{equation}{0}}

\def\one{{\hbox{ 1\kern-.8mm l}}}
\def\zero{{\hbox{ 0\kern-1.5mm 0}}}
 
\begin{document}
{}~
{}~
\hfill\vbox{\hbox{hep-th/0303057}%\hbox{MRI-P-020401}
}\break
 
\vskip .6cm
\centerline{\Large \bf Dirac-Born-Infeld Action on the}
\medskip
\centerline{\Large \bf  Tachyon Kink and 
Vortex}

\medskip
 
\vspace*{4.0ex}
 
\centerline{\large \rm
Ashoke Sen}
 
\vspace*{4.0ex}

\centerline{\large \it Harish-Chandra Research
Institute}

\centerline{\large \it  Chhatnag Road, Jhusi,
Allahabad 211019, INDIA}
 
\centerline{E-mail: ashoke.sen@cern.ch,
sen@mri.ernet.in}
 
\vspace*{5.0ex}
 
\centerline{\bf Abstract} \bigskip

The tachyon effective field theory describing the dynamics of a non-BPS
D-brane in superstring theory has an infinitely thin but finite tension
kink solution describing a codimension one BPS D-brane. We study the
world-volume theory of massless modes on the kink, and show that the world
volume action has precisely the Dirac-Born-Infeld (DBI) form without any
higher derivative corrections.  We generalize this to a vortex solution in
the effective field theory on a brane-antibrane pair. As in the case of
the kink, the vortex is infinitely thin, has finite energy density, and
the world-volume action on the vortex is again given exactly by the DBI
action on a BPS D-brane.  We also discuss the coupling of fermions and
restoration of supersymmetry and $\kappa$-symmetry on the world-volume of
the kink. Absence of higher derivative corrections to the DBI action on
the soliton implies that all such corrections are related to higher
derivative corrections to the original effective action on the
world-volume of a non-BPS D-brane or brane-antibrane pair.

\vfill \eject
 
\baselineskip=16.4pt

\sectiono{Introduction} \label{s1}

Study of various aspects of tachyon dynamics on a non-BPS D-brane of type
IIA or IIB superstring field theory has led to some understanding of the 
tachyon dynamics near the tachyon vacuum. 
The proposed tachyon effective action, describing the dynamics of the 
tachyon field on a non-BPS D$p$-brane of type IIA or IIB superstring 
theory, 
is given 
by\cite{9909062,0003122,0003221,0004106,0204143,0209122}:\footnote{Although 
we shall carry out our analysis for this 
action, our results are valid for a more general class of 
actions discussed in \cite{0209122,0011226,0012222,0208217}, where the 
lagrangian 
density in the absence of gauge and massless scalar fields takes the form 
$-V(T) \, 
F(\eta^{\mu\nu}\, \p_\mu T\p_\nu T)$, and $F(u) \sim 
u^{1/2}$ for large $u$. This follows from the fact that for the solutions 
we shall be considering, $u\equiv\eta^{\mu\nu}\p_\mu T\p_\nu T$ is large 
everywhere, and hence in this regime all these actions reduce to 
\refb{ez1}. Generalization 
of the action $-\int d^{p+1}x V(T) \,
F(\eta^{\mu\nu}\, \p_\mu T\p_\nu T)$
to include the world-volume gauge and scalar 
fields can be carried out by replacing $\eta^{\mu\nu}$ by the open string 
metric $G^{\mu\nu}$\cite{9903205,9908142}, and multiplying the action by 
an overall factor of 
$\sqrt{-\det(g_{\mu\nu}+F_{\mu\nu})}$, $g_{\mu\nu}=\eta_{\mu\nu}+\p_\mu 
Y^I \p_\nu Y^I$ being the induced closed string metric on the D-brane 
world-volume. This class of actions includes the action
proposed in \cite{0011226,0012222,0208217,0205085,0205098}, 
motivated by 
boundary 
string 
field theory\cite{0010108,0106231}. \label{f1}.}
\ben \label{ez1}
S &=& \int d^{p+1} x \, \LL\, , \nonumber \\
\LL &=& - V(T) \, \sqrt{-\det \bA}
\, ,
\een
where
\be \label{ey2}
\bA_{\mu\nu} = \eta_{\mu\nu} + \p_\mu T \p_\nu T + \p_\mu Y^I \p_\nu Y^I + 
F_{\mu\nu}\, ,
\ee
\be \label{ey2a}
F_{\mu\nu} = \p_\mu A_\nu - \p_\nu A_\mu\, .
\ee
$A_\mu$ and $Y^I$ for $0\le \mu, \nu\le p$, $(p+1)\le I\le 9$ are the 
gauge and the transverse scalar fields on the world-volume of the non-BPS 
brane, and $T$ is the tachyon field.
$V(T)$ is the tachyon potential which is symmetric under 
$T\to -T$, has a maximum at $T=0$, and has its minimum at $T=\pm\infty$ 
where it vanishes. We are 
using the convention where $\eta=diag(-1, 1, \ldots 1)$ and the 
fundamental string tension has been set equal to $(2\pi)^{-1}$ 
({\it i.e.} $\alpha'=1$).

The effective field theory described by the action \refb{ez1} is expected
to be a good description of the system under the condition that 1)  
$T$ is large, and 2) the second and higher derivatives of $T$ are small.
A kink solution in the full tachyon effective field theory, which is 
supposed to describe a BPS D-$(p-1)$-brane\cite{9808141,9812031,9812135},  
interpolates between the vacua at 
$T=\pm \infty$, and hence $T$
must pass through 0. Thus {\it a priori} we would expect that higher 
derivative 
corrections to the action \refb{ez1}
will be needed to provide a good description of the 
D-$(p-1)$-brane as a kink solution.
Nevertheless it is known 
that the energy density on the kink in the theory 
described by the action \refb{ez1} is 
localized strictly on a 
codimension one surface\cite{0011226,0012222,0208217,0301101} as in the 
case of a 
BPS
D-$(p-1)$-brane. We show that the world-volume theory on this kink 
solution 
is also
given precisely by the Dirac-Born-Infeld (DBI) action on a BPS 
D-$(p-1)$-brane. This agreement continues to hold even after including the 
world-volume fermion fields in the action, and we recover the expected 
supersymmetry and $\kappa$-symmetry on the BPS D-$(p-1)$-brane 
world-volume\cite{9610148,9610249,9611159,9611173,9612080}. Thus contrary 
to expectation, 
the kink solution of the effective field theory does provide a good 
description of the D-$(p-1)$-brane even without taking into account higher 
derivative corrections.

There have been several previous attempts to analyze the dynamics of 
fluctuations on a tachyon kink solution. 
Ref.\cite{0011226}
analyzed the world-volume theory of the
fluctuations on the kink. However in this study they restricted to the
analysis of small fluctuations.
We shall not put such restrictions, since in order to see the full DBI 
action we need to keep terms involving arbitrary power of the world-volume 
fields. 
Ref.\cite{0104218} analyzed the world-volume action on the kink keeping 
the non-linear terms, but including only the fluctuations in the 
transverse scalar field. 
Ref.\cite{0212188} addresses the problem of getting the DBI action on the 
soliton from the conformal field theory viewpoint, whereas 
ref.\cite{0102174} discusses construction of various special classical 
solutions of the 
tachyon effective field theory around the kink solution, without doing a 
general analysis of the equations of motion around this background.
A general approach to getting the DBI action on the kink and vortex 
solutions 
has
been described in \cite{0012080,0202079}. These papers, however, worked 
with 
very 
general form of the tachyon effective action, and arrived at the DBI 
action after ignoring the higher derivative terms.
In 
contrast,
we work with a specific form of the action given in \refb{ez1}, but given 
this form, 
we make no further
approximation in our 
analysis. In particular, we keep all powers of fields and all derivative 
terms, and nevertheless arrive at the DBI action without any higher 
derivative terms.
We should, of course, keep in mind that the action \refb{ez1} itself is at 
best an approximate 
action for the tachyon in string theory, and corrections to this action 
will certainly modify the world-volume action on the kink.
The significance of our result is that all such corrections involving 
higher derivative terms on the world-volume action 
of the BPS D-$(p-1)$-brane must come from explicit addition of such 
corrections to the world-volume action of the non-BPS D-$p$-brane. This 
suggests a sense in which the action \refb{ez1} is a `low energy effective 
action', -- namely that it reproduces the low energy effective 
action on the world-volume of the soliton without any correction terms.
In fact we also argue that in the world-volume theory on the kink, the 
would be massive modes, obtained by analysing the linearized equations of 
motion of various fields around the kink solution\cite{0011226}, disappear 
when we take 
into account the effect of the non-linear terms. Thus the only 
perturbative excitations on the kink world-volume are the massless degrees 
of freedom.

We also generalize our analysis to the construction of a vortex solution 
on 
a D$p$-brane - anti-D$p$-brane pair. For this we begin with a 
generalization 
of the tachyon effective action on brane-antibrane pair, -- this is done 
in a way that satisfies various known consistency requirements for such an 
action. We then construct the vortex solution, and find that it 
has finite energy density per unit 
$(p-2)$-volume; however the energy density is strictly localized on a 
codimension 2 subspace. Furthermore, the world-volume theory on the vortex 
is given by the DBI action expected for a BPS D-$(p-2)$-brane.

The rest of the paper is organized as follows. In section \ref{s2} we
review construction of the kink solution on a non-BPS D-brane.
In section \ref{s3} we analyze the world-volume 
theory of the bosonic fields on the kink.
In section \ref{s5a} we discuss the
coupling of fermions on the non-BPS and BPS D-brane world-volume, and show
how the supersymmetry and $\kappa$-symmetry, expected to be present on the 
world-volume action of
a BPS D-$(p-1)$-brane, appear in the world-volume action on the tachyon
kink in a non-BPS D-$p$-brane.
Section
\ref{s4} is devoted to construction of the vortex solution on the
brane-antibrane pair, and in section \ref{s5} we construct the
world-volume action on the vortex. 
We conclude with a few general comments
in section \ref{s6}.

\sectiono{The Kink Solution} \label{s2}

The construction of the kink solution follows 
\cite{0004106,0011226,0012222,0208217}.
The energy momentum tensor~\footnote{In writing 
down the
expression for the energy momentum
tensor, it will be understood that these are localized on the plane of the 
original
D-$p$-brane by a position space delta function in the transverse 
coordinates.
Also only the components of the energy-momentum tensor along the
world-volume of the original D-$p$-brane are non-zero.}
associated with the action \refb{ez1} is given 
by\cite{0009061,0204143,0208142}
\be \label{etmunu}
T^{\mu\nu} = - V(T) \, (\bA^{-1})_S^{\mu\nu} \, \sqrt{-\det \bA}\, ,
\ee
where the subscript $S$ denotes the symmetric part of a matrix.
In order to construct a kink solution, we look for a solution for which 
the tachyon
depends on one spatial direction $x\equiv x^p$ and is time independent, 
and furthermore, the gauge fields and the transverse scalar fields are set 
to zero. 
For such a background
the energy momentum tensor is given by:
\ben \label{ezsol1}
&& T_{xx} = -V(T) / \sqrt{1+(\p_x T)^2}\, , \qquad T_{\alpha x} = 0,
\nonumber \\
&& T_{\alpha\beta} = - V(T) \, \sqrt{1+(\p_x T)^2} \, \eta_{\alpha\beta}\, 
,
\quad
\hbox{for} \quad 0\le\alpha,\beta\le (p-1)\, .
\een
The energy-momentum conservation gives,
\be \label{ezsol2}
\p_x T_{xx} = 0\, .
\ee
Thus $T_{xx}$ is independent of $x$. Since for a kink solution
$T\to\pm\infty$ as $x\to\pm\infty$, and $V(T)\to 0$ in this limit, 
$T_{xx}$ vanishes as $x\to\infty$. Thus
$T_{xx}$ must vanish for all $x$. This, in turn, shows that we must
have
\be \label{ezsol3}
T=\pm\infty, \quad \hbox{or} \quad \p_x T= \infty \quad \hbox{(or
both)} \quad \hbox{for all $x$}\, .
\ee
Clearly the solution looks singular. We shall now see that despite this
singularity, the solution has finite energy density which is 
independent
of the way we regularize the singularity. Also the energy
density is localized on a
codimension 1 subspace, just as is expected of a 
D$(p-1)$-brane\cite{0011226,0208217}. For this
let us consider the field configuration
\be \label{ezsol4}
T(x) = f(ax)\, ,
\ee
where $f(u)$ satisfies
\be \label{efx}
f(-u) = - f(u), \qquad f'(u) > 0 \quad \forall \quad u, \qquad 
f(\pm\infty) = \pm\infty\, ,
\ee
but is otherwise an arbitrary function of its argument $u$.
$a$ is a constant that we shall take to $\infty$ at the end. In
this limit we have $T=\infty$ for $x>0$ and $T=-\infty$ for $x<0$. Thus 
the kink is
singular as expected.
Eq.\refb{ezsol1} gives the non-zero
components of $T_{\mu\nu}$ for this background to be:
\be \label{ezsol5}
T_{xx} = -V(f(ax))\Big/ \sqrt{1 + a^2 (f'(ax))^2}\, , \qquad
T_{\alpha\beta} = - V(f(ax))\, \sqrt{1 + a^2 (f'(ax))^2}\, 
\eta_{\alpha\beta}\, .
\ee
Clearly in the $a\to\infty$ limit, $T_{xx}$ vanishes everywhere since
the numerator vanishes (except at $x=0$) and the denominator blows up 
everywhere. Hence
the conservation law \refb{ezsol2} is automatically satisfied.

Let us now check that this configuration satisfies the full set of 
equations of motion. The non-trivial components of the equations of motion 
are:
\be \label{eeom}
\p_x \left({V(T) \p_x T\over \sqrt{1 + (\p_x T)^2}}\right) - V'(T) \, 
\sqrt{1+ (\p_x T)^2} = 0\, .
\ee
Taking $T=f(ax)$ we get the left hand side to be:
\ben \label{eeoma}
&& \p_x \left({V(f(ax)) a f'(ax) \over \sqrt{1 + a^2 (f'(ax))^2}}\right) - 
V'(f(ax)) 
\,\sqrt{1 + a^2 (f'(ax))^2} \nonumber \\
&=& V'(f(ax)) {a^2 (f'(ax))^2 \over 
\sqrt{1 + 
a^2 (f'(ax))^2}} + {V(f(ax)) a^2 f''(ax) \over 
(1 +
a^2 (f'(ax))^2)^{3/2}} - V'(f(ax))
\,\sqrt{1 + a^2 (f'(ax))^2} \nonumber \\
&=& \OO\left({1\over a}\right)\, ,
\een
assuming that $V(y) f''(y) / (f'(y))^3$ does not blow up anywhere.
Thus in the $a\to \infty$ limit the configuration satisfies the equations 
of motion. 

We shall now compute the energy-density associated with this 
solution. 
{}From \refb{ezsol5} we see  that in the $a\to\infty$ limit $T_{xx}$ 
vanishes, and we
can write $T_{\alpha\beta}$ as:
\be \label{ezsol6}
T_{\alpha\beta} = -a\, \eta_{\alpha\beta}\,  V(f(ax))\, f'(ax)\, .
\ee
Thus the integrated $T_{\alpha\beta}$, associated with the codimension 1
soliton, is given by:
\be \label{ezsol7}
T^{kink}_{\alpha\beta} = -a\, \eta_{\alpha\beta} \, 
\int_{-\infty}^\infty\, 
d x \,
V(f(ax))\, f'(ax) = -\eta_{\alpha\beta} \, \int_{-\infty}^\infty \, d y \,
V(y)\, ,
\ee
where $y=f(ax)$. Thus $T^{kink}_{\alpha\beta}$
depends only on the form of $V(y)$ and not on the shape of
the function $f(u)$ used to describe the soliton\cite{0208217,0102174}. It 
is also 
clear from
the exponential fall off in $V(y)$ for large $y$ that most of the
contribution to $T^{kink}_{\alpha\beta}$
is contained within a finite range of $y$.
{}From the relation $y=f(ax)$ we see that this
means that the contribution comes from a region of $x$ integral of
width $1/a$ around $x=0$. In the $a\to\infty$ limit such a distribution
approaches a $\delta$-function. Thus the 
$(p+1)$-dimensional energy-momentum tensor associated 
with
this solution is given by:
\be \label{ezsol8}
T_{xx}=0\, , \qquad T_{\alpha\beta} = - \eta_{\alpha\beta} \, \delta(x) \, 
\int_{-\infty}^\infty \, d y
\,
V(y)\, .
\ee
This is precisely what is expected of a D-$(p-1)$-brane, provided the
integral $\int_{-\infty}^\infty \, d y \, V(y)$ equals the tension of the
D-$(p-1)$-brane. 
For comparison, we also recall that $V(0)$ denotes the 
tension
of a D$p$-brane. These relations can be written 
as:\footnote{For more general actions of the kind discussed 
in footnote \ref{f1}, if $F$ is normalized such that $F(0)=1$, and if
for large $u$, $F(u)/u^{1/2} \simeq C$, then we have $\TT_p=V(0)$, 
$\TT_{p-1}=C\int_{-\infty}^\infty V(y) dy$. $V(T)$ and $F(u)$ motivated 
by boundary string field theory automatically gives the correct ratio of 
the D$p$-brane and D-$(p-1)$-brane tensions\cite{0011226}.} 
\be \label{ezsol9}
\TT_p = V(0), \qquad  \TT_{p-1} = \int_{-\infty}^\infty V(y) dy\, .
\ee
If we also require that the 
tachyon around 
$T=0$ has mass$^2=-{1\over 2}$, we 
get\cite{0011226}
$V''(0) / V(0) = -1/2$. However, higher derivative 
contribution to the action could modify this result.

Incidentally, we might note that one possible choice of the function 
$f(u)$ is $f(u)=u$. For this choice, the second and higher derivatives of 
the tachyon field vanish everywhere.\footnote{We note the similarity 
between such solutions and those in boundary 
superstring field theory\cite{0010108}. This of course is consistent 
with 
the proposal that the effective action from boundary string field theory 
has the general form given in footnote 
\ref{f1}\cite{0011226,0012222,0205085,0205098}.}
Thus the tachyon satisfies at least one of the 
two conditions under which the effective action \refb{ez1} is expected to 
be valid. 
The agreement between the properties of the soliton and those of 
a D-$(p-1)$-brane suggests that 
corrections to the action \refb{ez1} organize themselves in a way so as 
not to affect the desired features of the kink solution of \refb{ez1}.

We conclude this section by giving an intuitive argument for the infinite 
spatial gradient of $T$. From eq.\refb{ezsol1} we see that the total 
energy associated with a static configuration depending on only one 
spatial direction $x$, and interpolating between $T=\pm\infty$ at 
$x=\pm\infty$, is given by:
\ben \label{ebound}
E &=& \int_{-\infty}^\infty dx \, V(T(x)) \, \sqrt{1 + (\p_x T)^2} 
\nonumber \\
&\ge& 
\int_{-\infty}^\infty dx \,V(T(x)) |\p_x T| \ge \int_{-\infty}^\infty 
dx \,V(T(x)) 
\p_x T = \int_{-\infty}^\infty dy V(y) \, .
\een
The right hand side of \refb{ebound} is independent of the choice of 
$T(x)$. Since a static solution of the equations of motion must minimize 
(extremize) the total energy, we conclude that in order to get a solution 
of the equations of motion the bound given in \refb{ebound} must be 
saturated. This requires $|\p_x T|\to\infty$ and $\p_x T >0$ everywhere. 
This is precisely the result we obtained by explicitly analyzing the 
equations of motion.

\sectiono{Study of Fluctuations Around the Kink} \label{s3}

In this section we shall study fluctuations of various bosonic fields 
around the kink 
background and compare the effective action describing the dynamics of 
these fluctuations to the expected DBI action on the 
D-$(p-1)$-brane world-volume. 
First as a warm-up exercise we shall consider the dynamics of the 
translation zero mode along the $x$ direction, keeping the gauge fields 
$A_\mu$ and the transverse scalar fields $Y^I$ to zero. Such fluctuations 
correspond to fluctuation of $T$ 
of the form:
\be \label{ezsol10}
T(x, \xi) = f(a(x-t(\xi)))\, ,
\ee
where we have denoted by $\{\xi^\alpha\}$ for $0\le\alpha\le (p-1)$ the 
coordinates tangential to the kink world-volume. Here $t(\xi)$ is the 
$(p-1,1)$ dimensional field associated with the translational zero mode of 
the kink.\footnote{Since the soliton solution is infinitely thin, we do 
not need to rescale the argument of $f$ by $\sqrt{1 + \p^\alpha t\p_\alpha 
t}$ as in \cite{0012080}.} For this configuration,
\be \label{ezsol11}
-\det(\bA) = (1 + \eta^{\mu\nu} \p_\mu T \p_\nu T)
= 1 + a^2 (f')^2 (1 + \eta^{\alpha\beta} \p_\alpha t \p_\beta t)\, ,
\ee 
where for brevity we have denoted $f'(a(x-t(\xi)))$ by $f'$, and 
$f(a(x-t(\xi)))$ by $f$.
Substituting this into the action \refb{ez1} we get, for $a\to\infty$:
\be \label{ezsol12}
S = -\int d^p\xi \, \int dx\, V(f) \, a \, f' \, \sqrt{1 + 
\eta^{\alpha\beta} \p_\alpha t \p_\beta t}\, .
\ee
We now make a change of variables from $x$ to $y$:
\be \label{ezsol13}
y = f(a (x - t(\xi)))\, .
\ee
\refb{ezsol12} may then be rewritten as
\be \label{ezsol14}
S = -\int d^p\xi \, \int_{-\infty}^\infty \, dy \, V(y) \, \sqrt{1 +
\eta^{\alpha\beta} \p_\alpha t \p_\beta t}\, .
\ee
Performing the $y$
integral, and using \refb{ezsol9} we get
\be \label{ezsol15}
S = -\TT_{p-1} \int d^p\xi \,\sqrt{1 +
\eta^{\alpha\beta} \p_\alpha t \p_\beta t}\, .
\ee
This is precisely the action involving the scalar field $t$ associated 
with the coordinate $x$ transverse to a D-$(p-1)$-brane, lying in the 
$\xi^1,\ldots 
\xi^{p-1}$ plane. For the boundary string field theory action, this 
analysis was carried out previously in \cite{0104218}.

Note, however, that this does not yet establish that the dynamics of the
kink is described by the action \refb{ezsol15}. In 
order to do so, we need to establish that given any solution of the 
equations of motion derived from \refb{ezsol15}, it will produce a 
solution of 
the original equations of motion derived from the action \refb{ez1} under 
the identification \refb{ezsol10}.
Put another way, since $S$ given in \refb{ez1} reduces to that given in 
\refb{ezsol15} when \refb{ezsol10} holds, 
we already know that given a solution of the equations of motion 
of \refb{ezsol15},  $\delta S$ vanishes for any variation of $T$ that is 
induced due to a variation of $t(\xi)$ through \refb{ezsol10}. What needs 
to be shown is that $\delta S$ also vanishes for a 
$\delta T$ with more general $x$-dependence that is not necessarily 
induced due to a variation $\delta t(\xi)$ 
of $t$. For this we need to look at the general equation of motion of $T$ 
following from \refb{ez1}. It is:
\be \label{ezsol17}
\eta^{\alpha\beta} \p_\alpha \left({V(T) \p_\beta T\over 
\sqrt{1+\eta^{\mu\nu}\p_\mu T \p_\nu T}}\right)
+ \p_x \left({V(T) \p_x T \over \sqrt{1+\eta^{\mu\nu}\p_\mu T \p_\nu 
T}}\right)
- V'(T) \sqrt{1+\eta^{\mu\nu}\p_\mu T \p_\nu T} = 0\, .
\ee
Substituting \refb{ezsol10} into \refb{ezsol17}, and using the equations 
of motion of $t(\xi)$ derived from \refb{ezsol15} we can easily verify 
that the left hand side of \refb{ezsol17} vanishes in the $a\to\infty$ 
limit. This, in turn, shows that the dynamics of the field $t(\xi)$ is 
described precisely by the action \refb{ezsol15}.

Let us now turn to the inclusion of the gauge fields $A_i$ and the scalar 
fields $Y^I$. We expect that appropriate fluctuations in these fields will 
be responsible for the transverse scalar field excitations $y^I$ and gauge 
field 
excitations $a_\alpha$ on the D-$(p-1)$-brane. Thus the first step is to 
make 
a 
suitable ansatz for the fluctuations in the $(p+1)$-dimensional fields 
$A_\mu$ and $Y^I$ in 
terms of the $(p-1+1)$-dimensional fields $a_\alpha(\xi)$ and $y^I(\xi)$. 
We make the following ansatz:
\be \label{ezsol18}
A_x(x, \xi) = 0, \qquad A_\alpha(x, \xi) = a_\alpha(\xi), \qquad Y^I(x, 
\xi) = y^I(\xi)\, ,
\ee
together with \refb{ezsol10}.
In other words we take the fields $A_\mu$ and $Y^I$ to be independent of 
$x$. This seems surprising at first sight, since the fluctuations on a 
kink are expected to be localized around $x=0$ where the kink is sitting.
We note 
however that the 
dynamics of the gauge fields $A_\mu$ and the scalar fields $Y^I$ away from 
the 
location of the kink is essentially 
trivial\cite{9909062,0009038,0009061,0010240,0209034}, and 
hence 
although we allow 
fluctuations in $A_\mu$ and $Y^I$ far away from the location of the 
kink, the energy momentum tensor associated with such fluctuations is 
localized in the plane of the brane due to the explicit factor of $V(T)$ 
in \refb{etmunu} which vanishes away from the plane of the 
kink.\footnote{Only exceptions to this arises when the field strengths are 
at 
their critical values\cite{0002223,0009061,0010240}.} We shall discuss 
this issue further at the end of this section.

The next step will be to show that with the ansatz \refb{ezsol10}, 
\refb{ezsol18} the action \refb{ez1} reduces to the DBI action on a BPS 
D-$(p-1)$ brane. Computation of $\bA_{\mu\nu}$ defined in \refb{ey2} with 
this ansatz yields:
\ben \label{ezsol19}
&& \bA_{xx} = 1 + a^2 (f')^2, \qquad \bA_{x\alpha} = \bA_{\alpha x} = -a^2 
(f')^2 
\p_\alpha t, \nonumber \\
&& \bA_{\alpha\beta} = (a^2 (f')^2 -1) \p_\alpha t 
\p_\beta t + \AAA_{\alpha\beta}\, ,
\een
where $f\equiv f\big(a(x-t(\xi))\big)$, $f'\equiv 
f'\big(a(x-t(\xi))\big)$, and
\be \label{ezsol24}
\AAA_{\alpha\beta} = \eta_{\alpha\beta} + f_{\alpha\beta} + \p_\alpha
y^I \, \p_\beta y^I + \p_\alpha t \, \p_\beta t \, , \qquad 
f_{\alpha\beta} = \p_\alpha a_\beta - \p_\beta a_\alpha\, .
\ee
We can simplify the evaluation of $\det \bA$ by adding
appropriate 
multiples of the first row and first column to other rows and columns. 
More specifically, we define:
\ben \label{ezsol20}
&& \wh \bA_{\mu\beta} = \bA_{\mu\beta} + \bA_{\mu x} \p_\beta t, \qquad 
\wh 
\bA_{\mu 
x} = \bA_{\mu x}\, , \nonumber \\
&& \wt \bA_{\alpha\nu} = \wh \bA_{\alpha\nu} + \wh \bA_{x\nu} \p_\alpha t, 
\qquad \wt \bA_{x\nu} = \wh \bA_{x\nu}\, .
\een
Clearly this operation does not change the determinants; so we have
\be \label{ezsol21}
\det(\bA) = \det(\wh \bA) = \det(\wt \bA)\, .
\ee
On the other hand, we have, from \refb{ezsol19}, \refb{ezsol20}, 
\ben \label{ezsol22}
&& \wt \bA_{xx} = 1 + a^2 (f')^2, \qquad \wt \bA_{x\alpha}=\wt \bA_{\alpha 
x} = 
\p_\alpha t, \nonumber \\
&& \wt \bA_{\alpha\beta} = \AAA_{\alpha\beta}
\, .
\een
Using \refb{ezsol21}, \refb{ezsol22}, we get
\be \label{ezsol23}
\det (\bA) = a^2 (f')^2 [\det \AAA + \OO\left({1\over a^2}\right)]\, .
\ee
Substituting this into \refb{ez1}, we get, in the $a\to\infty$ limit,
\be \label{ezsol25}
S = - \int d^p\xi \, \int dx \, V(f) a f' \sqrt{-\det \AAA}
\, .
\ee
Making the change of variables \refb{ezsol13} and using \refb{ezsol9} we 
can write this as
\be \label{ezsol26}
S = - \TT_{p-1} \, \int d^p\xi \, \sqrt{-\det \AAA}\, ,
\ee
with $\AAA_{\alpha\beta}$ given by \refb{ezsol24}. This is precisely the 
world-volume 
action on a BPS D-$(p-1)$-brane if we identify the field $t$ as the 
coordinate $y^p$ associated with the $p$-th direction.

In order to establish that the dynamics of the kink is described by the
action \refb{ezsol26}, we now need to show that any solution of the
equations of motion derived from the action \refb{ezsol26} also provides a 
solution of the full $(p+1)$-dimensional equations of motion. The 
$p$-dimensional equations, derived from \refb{ezsol26} are:
\ben\label{ezsol27}
&& \p_\alpha \left( { (\AAA^{-1})^{\alpha\beta}_S \p_\beta t
\sqrt{-\det \AAA}}\right) = 0\, , \nonumber \\
&& \p_\alpha \left( { (\AAA^{-1})^{\alpha\beta}_S \p_\beta y^I
\sqrt{-\det \AAA}}\right) = 0\, , \nonumber \\
&& \p_\alpha \left( { (\AAA^{-1})^{\alpha\beta}_A
\sqrt{-\det \AAA}}\right) = 0\, ,
\een
where the subscripts $S$ and $A$ denote the symmetric and anti-symmetric 
components of a matrix respectively.
On the other hand the $(p+1)$-dimensional equations, which need to be 
verified, are
\ben\label{ezsol28} 
&& \p_\mu \left( {V(T)\, (\bA^{-1})^{\mu\nu}_S \p_\nu T \sqrt{-\det 
\bA}}\right) 
- 
V'(T) \sqrt{-\det \bA} = 0\, , \nonumber \\
&&\p_\mu \left( {V(T)\, (\bA^{-1})^{\mu\nu}_S \p_\nu Y^I \sqrt{-\det 
\bA}}\right) = 
0\, , \nonumber \\
&& \p_\mu \left( {V(T)\, (\bA^{-1})^{\mu\nu}_A \sqrt{-\det \bA}}\right) =
0\, .
\een
Eqs.\refb{ezsol10} and \refb{ezsol18} expresses the $(p+1)$-dimensional 
fields in terms of $p$-dimensional fields. We also need expressions for 
$\bA^{-1}$ and $\det(\bA)$ in terms of $\AAA_{\alpha\beta}$. These are 
summarized in the relations:
\ben \label{ezsol29}
&& (\bA^{-1})^{xx} \simeq (\AAA^{-1})^{\alpha\beta} \p_\alpha t \p_\beta t\, 
, 
\qquad
(\bA^{-1})^{x\alpha} \simeq \p_\beta t \, (\AAA^{-1})^{\beta\alpha}, 
\nonumber 
\\
&& (\bA^{-1})^{\alpha x} \simeq (\AAA^{-1})^{\alpha\beta}\, \p_\beta t\, , 
\qquad 
(\bA^{-1})^{\alpha\beta} \simeq  (\AAA^{-1})^{\alpha\beta}\, ,
\een
together with eq.\refb{ezsol23}. All the relations given in \refb{ezsol29} 
hold up to corrections of order $1/a^2$. 

We shall now verify that eqs.\refb{ezsol27}, together with 
\refb{ezsol10}, \refb{ezsol18}, implies eqs.\refb{ezsol28}. Besides the 
relations \refb{ezsol23}, \refb{ezsol29}, an identity that 
is 
particularly useful in carrying out this analysis is:
\be \label{ezsol30}
\p_\alpha F(x-t(\xi)) = -\p_\alpha t \, \p_x  F(x-t(\xi))\, ,
\ee
for any function $F$. 
We begin our discussion with the verification of the second 
equation of \refb{ezsol28}. 
Using eqs.\refb{ezsol10}, \refb{ezsol18}, \refb{ezsol23} and 
\refb{ezsol29} we can express the left hand side 
of this equation as:
\ben \label{ezsol31}
&& \p_x \{ V(T) (\bA^{-1})^{x\beta}_S \p_\beta Y^I \, \sqrt{-\det \bA} \}
+ \p_\alpha \{ V(T) (\bA^{-1})^{\alpha\beta}_S \p_\beta Y^I \, \sqrt{-\det 
\bA} \}
\nonumber \\
&& \simeq \p_x \{ V(f) \p_\alpha t \p_\beta y^I 
(\AAA^{-1})^{\alpha\beta}_S \, a f' 
\, \sqrt{-\det \AAA} \}
+ \p_\alpha \{V(f)\, (\AAA^{-1})^{\alpha\beta}_S \p_\beta y^I \, a f'
\, \sqrt{-\det \AAA} \} \nonumber \\
&& = (\AAA^{-1})^{\alpha\beta}_S \p_\beta y^I \, \sqrt{-\det \AAA}\, \{ 
\p_\alpha t \, \p_x (V(f)a f') + \p_\alpha (V(f)a f')\} = 0\, ,
\een
where in going from the second to the third line we have used the second 
equation in
\refb{ezsol27}, and
in the last step we have used eq.\refb{ezsol30}.
Note however that only terms of order $a^2$ and $a$ cancel, leaving behind 
a contribution of order 1. These finite contributions come, for example, 
from product of $\OO(a^{-2})$ corrections to the right hand side of 
eqs.\refb{ezsol23},
\refb{ezsol29} with the $\OO(a^2)$ contribution from $\p_x(V(f) af')$. 
However, since $V(T)$ is non-zero only within a range of order $1/a$ in 
the 
$x$ space, the contribution to a variation $\delta S$ in the action due to 
the finite terms in the equations of motion will be of order $1/a$ for any 
finite $\delta Y^I$. This goes to zero in the $a\to\infty$ limit, and 
hence we conclude that the $y^I$ equations of motion given in 
\refb{ezsol27} implies $\delta S=0$ for arbitrary finite $\delta Y^I$.

Verification of the third equation of \refb{ezsol28} proceeds in the same 
way. For $\nu=\beta$ the left hand side of this equation is given by:
\ben \label{ezsol32}
&& \p_x \{ V(T) (\bA^{-1})^{x\beta}_A \, \sqrt{-\det \bA} \}
+ \p_\alpha \{ V(T) (\bA^{-1})^{\alpha\beta}_A \, \sqrt{-\det 
\bA} \}
\nonumber \\
&& \simeq \p_x \{ V(f) \p_\alpha t  
(\AAA^{-1})^{\alpha\beta}_A \, a f' 
\, \sqrt{-\det \AAA} \}
+ \p_\alpha \{V(f)\, (\AAA^{-1})^{\alpha\beta}_A \, a f'
\, \sqrt{-\det \AAA} \} \nonumber \\
&& = (\AAA^{-1})^{\alpha\beta}_A \, \sqrt{-\det \AAA} \, \{ 
\p_\alpha t \, \p_x (V(f)a f') + \p_\alpha (V(f)a f')\} = 0\, .
\een
In going from the second to the third line in \refb{ezsol32} we have used 
the last equation in \refb{ezsol27}.
Again \refb{ezsol32} has finite left-over contribution, but this is 
sufficient to establish that the variation of $\delta S$ vanishes for 
arbitrary finite $\delta A_\alpha$ when the equations \refb{ezsol27} are 
satisfied. 

For $\nu=x$, the left hand side of the third equation in 
\refb{ezsol28} has the form:
\ben \label{eax}
&& \p_\alpha \{ V(T) (\bA^{-1})^{\alpha x}_A \, \sqrt{-\det 
\bA} \}
\nonumber \\
&\simeq& \p_\alpha \{V(f)\, (\AAA^{-1})^{\alpha\beta}_A \p_\beta t \, a 
f'
\, \sqrt{-\det \AAA} \} \nonumber \\
&=& (\AAA^{-1})^{\alpha\beta}_A \, \p_\beta t \, \sqrt{-\det \AAA} \, 
\, \p_\alpha (V(f)a f') =
- (\AAA^{-1})^{\alpha\beta}_A \, \sqrt{-\det \AAA} \, \, \p_\beta
t \p_\alpha  t 
\p_x (V(f)a f') \nonumber \\
&=& 0\, ,
\een
where in going from the second to the third line of \refb{eax} we have 
used the third equation in \refb{ezsol27} and the antisymmetry of 
$(\AAA^{-1})^{\alpha\beta}_A$, and in the last step we 
have used the antisymmetry of $(\AAA^{-1})^{\alpha\beta}_A$.

Verification of the first equation of \refb{ezsol28} is a little more 
involved due to the following reasons. First of all, here the leading 
contribution from individual terms is of order $a^3$, with one factor of 
$a$ coming from $\sqrt{-\det \bA}$ and two more factors of $a$ coming from 
the two derivatives of $f\Big(a(x-t(\xi))\Big)$. Thus we cannot, from the 
beginning, use \refb{ezsol23} and \refb{ezsol29}, since the corrections of 
order $a^{-2}$ in 
these equations
could 
combine with the $a^3$ terms to give a contribution of order $a$. 
Furthermore, since finite $\delta t$ induces a $\delta T=-a f' \delta 
t\sim 
a$, the equations of motion of $T$ must hold including finite terms, since 
such terms will give a contribution of order 1 in $\delta S$. 
We proceed with our analysis as follows.
Using the equations 
$(\bA^{-1})^{\mu\nu} \bA_{\nu x} = 
\delta^\mu_x$, $\bA_{x\nu}(\bA^{-1})^{\nu\mu} = \delta^\mu_x$, we get the 
following exact 
relations:
\be \label{enew2}
(\bA^{-1})^{\mu x}_S - (\bA^{-1})^{\mu\beta}_S \p_\beta t = {1\over a^2 
(f')^2} \Big(\delta^\mu_x - (\bA^{-1})^{\mu x}_S\Big)\, .
\ee
Using \refb{enew2} and that $\p_\beta T = -\p_x T \p_\beta 
t=-af'\p_\beta t$, we can now 
express the left hand side ($l.h.s.$) of the first equation of 
\refb{ezsol28} as
\be \label{eteqn}
l.h.s. = \p_\mu \left( V(f) \, af' \, {1\over a^2 (f')^2} (\delta^\mu_x - 
(\bA^{-1})_S^{\mu x}) \sqrt{-\det \bA}\right) - V'(f) \sqrt{-\det \bA} \, 
.
\ee
Due to the explicit factor of $a^2 (f')^2$ in the denominator of the first 
term, the leading contribution from individual terms in this expression is 
now of order $a$, and 
hence we can now use eqs.\refb{ezsol23}, \refb{ezsol29} to analyze 
\refb{eteqn} if we are willing to ignore contributions of order $1/a$. 
Using these equations
\refb{eteqn} can be simplified as follows:
\ben \label{eteqn1}
l.h.s. &\simeq& \p_x \left\{ V(f) \, 
\sqrt{-\det 
\AAA} \, (1 - (\AAA^{-1})^{\alpha\beta}_S \p_\alpha t \p_\beta t)\right\}
\nonumber \\
&& - \p_\alpha\left\{ V(f) \, \sqrt{-\det
\AAA}\, \, (\AAA^{-1})^{\alpha\beta}_S \p_\beta t\right\} - V'(f) \, a f' 
\sqrt{-\det
\AAA} \nonumber \\
&=& V'(f)  a f'\sqrt{-\det
\AAA} \, \,  (1 - (\AAA^{-1})^{\alpha\beta}_S \p_\alpha t \p_\beta t)
+ V'(f)  a f' \, \sqrt{-\det
\AAA} \, \,  (\AAA^{-1})^{\alpha\beta}_S \p_\alpha t \p_\beta t \nonumber 
\\
&& - V(f) \p_\alpha \Big( \sqrt{-\det
\AAA} \, \, (\AAA^{-1})^{\alpha\beta}_S \p_\beta t\Big) - V'(f) a f' 
\sqrt{-\det
\AAA} \nonumber \\
&=& 0\, ,
\een
using the first equation of \refb{ezsol27}. This establishes that any 
solution of eqs.\refb{ezsol27} automatically gives a solution of 
eqs.\refb{ezsol28}. 

Before concluding this section, let us note that if 
we consider a general expansion of the fields $Y^I$ and $A_\mu$ of the 
form:
\ben \label{egen1}
&& Y^I(x,\xi) = y^I(\xi) + \sum_{n=1}^\infty f_n(x-t(\xi)) y^I_{(n)}(\xi), 
\nonumber 
\\
&& A_x(x,\xi) = \phi_{(0)}(\xi) + \sum_{n=1}^\infty f_n(x-t(\xi))
\phi_{(n)}(\xi) \equiv \phi(x, \xi)\, , \qquad 
\nonumber \\
&& A_\alpha(x,\xi) = a_\alpha(\xi) + \sum_{n=1}^\infty
f_n(x-t(\xi)) 
a^{(n)}_\alpha(\xi)- \phi(x, \xi) \p_\alpha t , 
\een
where $\{f_n(u)\}$ for $n\ge 1$ is a basis of smooth functions which 
vanish at $u=0$, 
and which are
bounded including at  
$u=\pm\infty$,\footnote{This condition is imposed so that $-\det(\bA)$ 
remains positive for all $x$ for arbitrary 
finite amplitude fluctuations of $y^I_{(n)}$, $a^{(n)}_\alpha$ 
and $\phi_{(n)}$.}
then the action will be independent of $y^I_{(n)}(\xi)$, 
$a^{(n)}_\alpha(\xi)$ for $n\ge 1$ and $\phi_{(n)}(\xi)$ for $n\ge 0$. 
This can be seen by carrying out the same manipulations on the matrix 
$\bA_{\mu\nu}$ as given in eqs.\refb{ezsol19}-\refb{ezsol26}.
This has the following implication.
As was argued in \cite{0009038,0011009}, at the tachyon vacuum a
finite deformation of the $A_\mu$ and the $Y^I$ fields do not change the
action, and hence it is natural to identify all such field configurations
as a single point in the configuration space, just like in the polar
coordinate system different values of the polar angle $\theta$ give rise
to the same physical point at $r=0$. This can be made into a general
principle by postulating that whenever we encounter a local transformation
that does not change the action, we should identify the different points
in the configuration space related by this local transformation. 
In this spirit, the deformations associated 
with $\phi(x,\xi)$, 
$y^I_{(n)}(\xi)$ and
$a^{(n)}_\alpha(\xi)$ should be regarded as pure gauge deformations.
This general principle
means, however, that the dimension of the gauge group may change
from one point to another in the configuration space, {\it e.g.} while 
around
the tachyon background all deformations in $A_\mu$ and $Y^I$ are pure
gauge, around the non-BPS D-$p$-brane solution most of these deformations
are physical, while around a kink solution some of these deformations are
physical. This should not come as a surprise, as it simply 
indicates
that the coordinate system that we have chosen, -- the fields $T$,
$A_\mu$ and $Y^I$, -- are not good coordinates everywhere in the
configuration space just like the polar coordiante system is not a good
system near the origin.

To summarize, what we see from this analysis is that not only is the 
effective field theory of low energy modes on the world-volume of the kink 
described by 
DBI action, but all the other smooth excitations on the kink world-volume 
associated with gauge and transverse scalar fields are pure gauge 
deformations. 
The action depends only on the pull-back of the fields $Y^I$ 
and the gauge field strength $F_{\mu\nu}$ along the surface $x=t(\xi)$ 
along which the kink world-volume lies. 
In particular, invariance of 
the action under the deformations generated by $y^I_{(n)}$, 
$a^{(n)}_\alpha$ and $\phi_{(n)}$ for $n\ge 1$ reflects that the action 
does not depend on the fields away from the location of the kink, 
whereas $\phi_{(0)}$ independence 
of the action reflects that the action depends only on the components of 
the gauge field strength along the world-volume of the kink.

In this context we would like to note that ref.\cite{0011226} analyzed the 
non-zero mode excitations involving the $A_\mu$ (and the tachyon) fields 
and found a non-trivial spectrum for these modes by working to quadratic 
order in these fields in the action. For potential $V(T)$ motivated by the 
boundary string field theory analysis, these eigenmodes turned out to be 
Hermite 
polynomials with their arguments scaled by $a$. Since these are not smooth 
functions in the $a\to 0$ limit, and blow up for large $x$ except for the 
constant mode, there is no conflict with our result. However we should 
note 
that in general, for actions of the kind considered here where the overall 
multiplicative factor vanishes away from the core of the soliton, the 
results based on the linearized analysis of the equations of motion may be 
somewhat misleading, since 
the non-linear terms could dominate even for small amplitude 
oscillations. In particular, if we consider the fluctuation of a mode of 
$A_\mu$ associated with a Hermite polynomial that grows for large $x$, 
then for any small but finite amplitude oscillation the $F_{\mu\nu}$ in 
$\bA_{\mu\nu}$ will become comparable with $\eta_{\mu\nu}$ for 
sufficiently large $x$, and could drive $-\det(\bA)$ to be negative, 
thereby invalidating the analysis. 
We can see this explicitly by taking the linear tachyon profile $T\propto 
ax$ as in \cite{0011226} and considering a fluctuation in the gauge field 
$A_1(x,\xi)$ of the form $H_n(ax) a_1(\xi^0)$ where $H_n$ denotes the 
$n$th Hermite polynomial. Let us further consider a specific instant of 
time when $a_1(\xi^0)$ vanishes but $\p_0 a_1(\xi^0)$ is non-zero. As this 
instant $\sqrt{-\det(\bA)}\propto a \sqrt{1 - (H_n(ax))^2 (\p_0 a_1)^2}$. 
Since $H_n(ax)$ grows for large $ax$, we see that for any finite $\p_0 
a_1$, however small, the expression under the square root will become 
negative for 
sufficiently large $ax$. The 
only mode that does not suffer from this problem is the constant mode. 
A similar argument holds for fluctuations in $Y^I$ and $T$.
This leads us to 
suspect that the only surviving modes on the kink world-volume are the 
massless modes associated with $t$, $y^I$ and $a_\alpha$.\footnote{This 
argument of course does not affect the analysis for other types of action
discussed in\cite{0009246,0011226} where the action takes the 
form of a kinetic plus a singular potential term.}
A similar argument works for potentials $V(T)$ with different asymptotic 
behaviour, {\it e.g.} $V(T) \sim e^{-\beta T}$ for large $T$ where
$\beta$ is some constant.
The only 
difference is that instead of the Hermite polynomials
$H_n(ax)$, we have some other 
functions which grow for large $ax$.

A simpler version of this problem can be seen even for studying gauge (and 
scalar) field fluctuations around the tachyon vacuum. If we expand the 
action $-C\int d^{p+1}x\,\sqrt{-\det (\eta+F)}$ to quadratic order in $F$, 
then we can 
absorb a factor of $\sqrt C$ in $A_\mu$ and get the standard kinetic term 
for the gauge fields. This would lead to a conclusion that the spectrum 
contains a massless photon for all $C$. However in the $C\to 0$ limit 
(relevant for the tachyon vacuum) this procedure is clearly incorrect 
since this will give an action $-C\int d^{p+1}x\,\sqrt{-\det (\eta+ 
C^{-{1\over 2}}F)}$, 
and even a small fluctuation in $F$ could drive the term under the 
square root negative, invalidating the analysis. In this case a 
Hamiltonian 
analysis of the system gives a much better understanding of the possible 
fluctuations around the tachyon vacuum\cite{0009061} (see also 
\cite{LIND}). A similar analysis 
in the kink background may provide useful insight into what type of 
fluctuations are present around this background.

\sectiono{World-volume Fermions, Supersymmetry and $\kappa$-symmetry}
\label{s5a}

So far in our discussion we have ignored the world-volume fermions. We
shall now discuss inclusion of these fields in our analysis.
 
For definiteness we shall restrict our analysis to D-branes in type IIA
string theory, but generalization to type IIB theory is straightforward 
following the analysis of ref.\cite{9909062}.
On a non-BPS D$p$-brane world-volume in type IIA string theory, we have a
32 component anti-commuting field $\Theta$ which transforms as a Majorana 
spinor of the 10 dimensional Lorentz group\cite{9909062}.
We shall denote by $\Gamma_M$ the ten dimensional $\gamma$-matrices, and
take
the indices $M,N$ to run from 0 to 9.
In order to construct the world-volume
action involving the
fields $A_\mu$, $Y^I$, $\Theta$ and $T$ ($0\le\mu\le p$, $(p+1)\le I\le
9$) in static gauge, we first define:
\be \label{efer1}
\Pi^\nu_\mu = \delta^\nu_\mu - \bar\Theta \Gamma^\nu \p_\mu \Theta, \qquad
\Pi^I_\mu = \p_\mu Y^I - \bar\Theta \Gamma^I \p_\mu \Theta,
\ee
\be \label{emissing}
\bG_{\mu\nu} = \eta_{MN} \Pi^M_\mu \Pi^N_\nu + \p_\mu T \p_\nu T\, ,
\ee
and
\be \label{efer3}
\bF_{\mu\nu} = F_{\mu\nu} - \Big[\{\bar\Theta \Gamma_{11} \Gamma_\nu
\p_\mu\Theta + \bar\Theta \Gamma_{11} \Gamma_I \p_\mu\Theta \p_\nu Y^I
- {1\over 2} \bar\Theta \Gamma_{11} \Gamma_M \p_\mu\Theta \bar\Theta
\Gamma^M \p_\nu\Theta\} -\{\mu\leftrightarrow\nu\}\Big]\, ,
\ee
where
\be \label{efer4}
F_{\mu\nu} = \p_\mu A_\nu - \p_\nu A_\mu\, .
\ee
In terms of these variables, the DBI part of the world-volume action
is given by\cite{9909062,0003122,0003221}:
\be \label{efer5}
S_{DBI} = -\int d^{p+1} x \, V(T) \, \sqrt{-\det(\bG+\bF)}\, .
\ee
The action is invariant under the supersymmetry transformation
parametrized by a ten dimensional Majorana spinor $\epsilon$. In the
static gauge in which we are working, the infinitesimal supersymmetry 
transformation
laws are given by\cite{9909062}:
\ben \label{efer6}
&&\delta_p\Theta = \eps - (\bar\eps\Gamma^\mu \Theta) \p_\mu\Theta, \qquad
\delta_p Y^I = \bar\eps \Gamma^I\Theta - (\bar\eps\Gamma^\mu \Theta)\p_\mu
Y^I\, ,\qquad \delta_p T =  - (\bar\eps\Gamma^\mu \Theta)\p_\mu T\, ,
\nonumber \\
&& \delta_p A_\nu = \bar\eps \Gamma_{11}\Gamma_\nu\Theta + \bar\eps
\Gamma_{11}\Gamma_I \Theta \p_\nu Y^I -{1\over 6} (\bar\eps
\Gamma_{11}\Gamma_M \Theta\, \bar\Theta \Gamma^M \p_\nu\Theta +
\bar\eps\Gamma_M \Theta\,\bar\Theta\Gamma_{11}\Gamma^M \p_\nu\Theta)
\nonumber \\
&& ~ \qquad \qquad \qquad - (\bar\eps\Gamma^\mu \Theta)\p_\mu A_\nu
-\p_\nu (\bar\eps\Gamma^\mu \Theta) A_\mu\, .
\een
The subscript $p$ in $\delta_p$ denotes that these are the supersymmetry 
transformation laws on the D-$p$-brane world-volume. The supersymmetry 
transformation parameter $\epsilon$ is a Majorana spinor of the ten 
dimensional Lorentz group. 

Besides the DBI term, the world-volume action also contains a Wess-Zumino
term. In the bosonic sector this term is important only for non-vanishing
RR background field, but once we take into account the world-volume
fermions, this term survives even for zero RR
background. The structure of
this term is\cite{9810188,9812135,9904207,0003221}:
\be \label{efer7}
S_{WZ}=\int W(T)\,  dT \wedge \bC \wedge e^\bF\, ,
\ee
where $\bF=\bF_{\mu\nu} dx^\mu\wedge dx^\nu$, 
$W(T)$ is an even function of $T$ which vanishes as $T\to\pm\infty$, 
and  $\bC$ is a specific
combination of background RR fields and the
world-volume fields $Y^I$, $\Theta$ on the D-brane\cite{0003221}.
In particular,
the bosonic part of
$\bC$ is given by
$\sum_{q\ge 0} C^{(p-2q)}$ where $C^{(p-2q)}$ denotes the pull-back of
the RR $(p-2q)$-form
field on the D-$p$-brane world-volume.
This vanishes for vanishing RR background, but there is a part of $\bC$ 
involving
the world-volume fermion fields that survives even in the absence of any 
RR
background\cite{9610148,9610249,9611159,9611173,9612080,0003221}.
Since we shall not need the explicit form of $\bC$ for our analysis, we 
shall not give it here.
(See, for example \cite{0003221} for the component form of this term for 
trivial supergravity background.)
The Wess-Zumino term is also invariant under the
supersymmetry transformations \refb{efer6}. Later we shall see that
consistency requires:
\be \label{efer8}
\int_{-\infty}^\infty W(T) dT = \int_{-\infty}^\infty V(T) dT =
\TT_{p-1}\, ,
\ee
where in the last step we have used eq.\refb{ezsol9}. 

Since we want to compare the world-volume action on a kink solution with
that on the BPS D-$(p-1)$-brane, we need to first know the form of the
world-volume action on a BPS D-$(p-1)$-brane. The world-volume fields in
this case consist of a vector field $a_\alpha(\xi)$ ($0\le\alpha\le 
(p-1)$), a set of $(9-p+1)$
scalar
fields which we shall denote by $y^I(\xi)$ ($(p+1)\le I\le 9$) and
$y^p(\xi)\equiv t(\xi)$
respectively
in the convention of section \ref{s3}, and a Majorana spinor $\theta(\xi)$
of
the ten dimensional Lorentz group.
Here $\{\xi^\alpha\}$ denote the world-volume
coordinate on the D-$(p-1)$-brane as in section \ref{s3}.
The DBI part of the action is given
by\cite{9610148,9610249,9611159,9611173,9612080}:
\be \label{efer15}
S_{dbi} = -\TT_{p-1}\, \int d^{p} \xi \, \sqrt{-\det(\ggg+\fff)}\, ,
\ee
where
\be \label{efer12}
\ggg_{\alpha\beta} = \eta_{MN} \pi^M_\alpha \pi^N_\beta
\, ,
\ee
\be \label{efer11}
\pi^\beta_\alpha = \delta^\beta_\alpha - \bar\theta \Gamma^\beta \p_\alpha
\theta,
\qquad
\pi^I_\alpha = \p_\alpha y^I - \bar\theta \Gamma^I \p_\alpha \theta,
\qquad \pi^p_\alpha = \p_\alpha t - \bar\theta \Gamma^p \p_\alpha
\theta,
\ee
\be \label{efer13}
\fff_{\alpha\beta} = f_{\alpha\beta} - \Big[\{\bar\theta \Gamma_{11}
\Gamma_\beta
\p_\alpha\theta + \bar\theta \Gamma_{11} \Gamma_I \p_\alpha\theta \p_\beta
y^I + \bar\theta \Gamma_{11} \Gamma_p \p_\alpha\theta \p_\beta
t
- {1\over 2} \bar\theta \Gamma_{11} \Gamma_M \p_\alpha\theta \bar\theta
\Gamma^M \p_\beta\theta\} -\{\alpha\leftrightarrow\beta\}\Big]\, ,
\ee
\be \label{efer14}
f_{\alpha\beta} = \p_\alpha a_\beta - \p_\beta a_\alpha\, .
\ee
The Wess-Zumino term, on the other hand, has the form:
\be \label{efer17}
S_{wz}=\TT_{p-1}\, \int \ccc \wedge e^{\fff}\, ,
\ee
where $\fff = \fff_{\alpha\beta} d\xi^\alpha\wedge d\xi^\beta$, and 
$\ccc$ is an expression containing the RR
background and the world-volume
fields $y^I,t,\theta$ \cite{9610148,9610249,9611159,9611173,9612080}.
The bosonic part of $\ccc$ is given by
$\sum_{q\ge 0} C^{(p-2q)}$ where $C^{(p-2q)}$ denotes the pull-back of
the RR $(p-2q)$-form
field on the D-$(p-1)$-brane world-volume. 
Like $\bC$, $\ccc$ also contains a term involving $y^I$ and $\theta$ which 
survive even for trivial RR background.
If we think of the world-volume
of the D-$(p-1)$-brane as sitting inside that of a D-$p$-brane along the 
surface $x^p=t(\xi)$, then 
$\ccc$ is in fact the pullback of $\bC$ appearing in \refb{efer7} provided 
we identify $\theta$ and $y^I$ as the restriction of $\Theta$ and $Y^I$ 
along the surface $x^p=t(\xi)$.

Both $S_{dbi}$ and $S_{wz}$ are separately invariant under the
infinitesimal supersymmetry transformation:
\ben \label{efer16}
&&\delta_{p-1}\theta = \eps - (\bar\eps\Gamma^\alpha \theta) 
\p_\alpha\theta,
\qquad
\delta_{p-1} y^I = \bar\eps \Gamma^I\theta - (\bar\eps\Gamma^\alpha
\theta)\p_\alpha
y^I\, ,\qquad \delta_{p-1} t = \bar\eps \Gamma^p\theta -
(\bar\eps\Gamma^\alpha
\theta)\p_\alpha t\, ,
\nonumber \\
&& \delta_{p-1} a_\beta = \bar\eps \Gamma_{11}\Gamma_\beta\theta + 
\bar\eps
\Gamma_{11}\Gamma_I \theta \p_\beta y^I
+ \bar\eps
\Gamma_{11}\Gamma_p\theta \p_\beta t
-{1\over 6} (\bar\eps
\Gamma_{11}\Gamma_M \theta\, \bar\theta \Gamma^M \p_\beta\theta +
\bar\eps\Gamma_M \theta\,\bar\theta\Gamma_{11}\Gamma^M \p_\beta\theta)
\nonumber \\
&& ~ \qquad \qquad \qquad - (\bar\eps\Gamma^\alpha \theta)\p_\alpha
a_\beta
-\p_\beta (\bar\eps\Gamma^\alpha \theta) a_\alpha\, .
\een
The subscript $(p-1)$ on $\delta_{p-1}$ indicates that these represent 
supersymmetry transformation laws on the world-volume of a BPS 
D-$(p-1)$-brane.

In order to show that the world-volume action $S_{dbi}+S_{wz}$ on
the BPS D-$(p-1)$-brane arises from the world-volume action on the tachyon
kink solution of section \ref{s2}, we need to first propose an ansatz
relating the fields $T(x,\xi)$, $A_\mu(x,\xi)$, $Y^I(x,\xi)$ and
$\Theta(x,\xi)$ to the fields $a_\alpha(\xi)$, $y^I(\xi)$,
$t(\xi)$ and $\theta(\xi)$ on the BPS D-brane. For this we propose the
following ansatz:
\ben \label{efer21}
&& T(x,\xi) = f\Big(a(x-t(\xi))\Big), \qquad Y^I(x,\xi) = y^I(\xi), \qquad
\Theta(x,\xi) = \theta(\xi), \nonumber \\
&& A_x(x,\xi) = 0 \qquad
A_\alpha(x,\xi) = a_\alpha(\xi) \, . \nonumber \\
\een
We can now compute $\bG_{\mu\nu}$ and $\bF_{\mu\nu}$ in terms of the 
variables $a_\alpha$, $y^I$, $t$ and $\theta$ using 
eqs.\refb{efer1}-\refb{efer4} and \refb{efer21}. The result is:
\ben \label{efer31}
&& \bG_{xx} = 1 + a^2 (f')^2\, , \qquad \bG_{\alpha x} = \bG_{x \alpha} = 
-a^2 (f')^2 \p_\alpha t - \bar\theta \Gamma^p \p_\alpha\theta\, , 
\nonumber 
\\
&& \bG_{\alpha\beta} = \ggg_{\alpha\beta} + \p_\alpha t \, \bar\theta 
\Gamma^p 
\p_\beta\theta + \p_\beta t \, \bar\theta \Gamma^p
\p_\alpha\theta + (a^2 (f')^2-1) \p_\alpha t \p_\beta t\, , \nonumber \\
&& \bF_{\alpha x} = -\bF_{x \alpha} = - \bar\theta \Gamma_{11} \Gamma^p 
\p_\alpha\theta\, , \nonumber \\
&& \bF_{\alpha\beta} = \fff_{\alpha\beta} - \p_\alpha t \, \bar\theta
\Gamma_{11} \Gamma^p
\p_\beta\theta + \p_\beta t \, \bar\theta \Gamma_{11} \Gamma^p
\p_\alpha\theta\, ,
\een
with $\ggg_{\alpha\beta}$ and $\fff_{\alpha\beta}$ defined as in 
eqs.\refb{efer12}-\refb{efer14}.
Using manipulations similar to those in eqs.\refb{ezsol20}-\refb{ezsol26}
we can now show 
that
\be \label{efer32}
\det(\bG+\bF) = a^2 (f')^2 \{ \det(\ggg+\fff) + \OO(a^{-2})\}\, ,
\ee
and 
\be \label{efer33}
S_{DBI} = -\int d^{p+1} x \, V(T) \, \sqrt{-\det(\bG+\bF)} = -\TT_{p-1} \, 
\int d^p \xi \, \sqrt{-\det(\ggg+\fff)} = S_{dbi}\, .
\ee

The analysis for $S_{WZ}$ is even simpler; -- indeed this term was 
designed to reproduce the Wess-Zumino term on the world-volume of a kink 
solution\cite{9812135,0003221}. For 
this let us define \be \label{efer41}
u = x - t(\xi)\, .
\ee
Then from \refb{efer31} we get
\ben \label{efer42}
\bF \equiv \bF_{\mu\nu}\, dx^\mu\wedge dx^\nu &=& 2 \bF_{x\beta}\, 
dx\wedge 
d\xi^\beta + \bF_{\alpha\beta}\, d\xi^\alpha\wedge d\xi^\beta\nonumber \\
&=& 2 
\bar\theta \Gamma_{11} \Gamma_p 
\p_\alpha\theta du \wedge d\xi^\alpha + \fff_{\alpha\beta}\, d\xi^\alpha 
\wedge 
d\xi^\beta\, .
\een
Since we have 
\be \label{efer43}
dT= a f'(au)\, du\, ,
\ee
only the second term on the right hand side of
\refb{efer42} will contribute to $S_{WZ}$ given in \refb{efer7}.
Thus we can replace $\bF$ by $\fff$ in \refb{efer7}. On the 
other hand, we can analyze $\bC$ by writing it as
\ben \label{efer44}
\bC &=& \sum_q \bC^{(q)}_{\mu_1\cdots \mu_q} dx^{\mu_1} \wedge \cdots 
dx^{\mu_q} \nonumber \\
& = & 
\sum_q (q\, \bC^{(q)}_{x \alpha_2\cdots \alpha_q} dx\wedge d\xi^{\alpha_2} 
\wedge 
\cdots 
d\xi^{\alpha_q} + \bC^{(q)}_{\alpha_1\cdots \alpha_q}d\xi^{\alpha_1} 
\wedge \cdots
d\xi^{\alpha_q}) \nonumber \\
&=& \sum_q\left[ q \, \bC^{(q)}_{x \alpha_2\cdots \alpha_q} du\wedge 
d\xi^{\alpha_2} \wedge \cdots
d\xi^{\alpha_q} + (q \, \bC^{(q)}_{x \alpha_2\cdots \alpha_q} 
\p_{\alpha_1} t + 
\bC^{(q)}_{\alpha_1\cdots \alpha_q})\, d\xi^{\alpha_1}\wedge \cdots
d\xi^{\alpha_q}\right]\, , \nonumber \\
\een
where in the last step we have used $dx = du + \p_\alpha t d\xi^\alpha$. 
The term proportional to $du$ does not contribute to \refb{efer7} due to 
eq.\refb{efer43}, whereas 
the term proportional to $d\xi^{\alpha_1} \wedge \cdots
d\xi^{\alpha_q}$, after being summer over $q$, is precisely the pull-back 
of $\bC$ on the 
kink world-volume along $x=t(\xi)$ and hence can be identified with 
$\ccc$. Thus we get
\be \label{efer45}
S_{WZ} = \int W(f(au)) \, a\, f'(au) \, du \wedge \ccc\wedge e^{\fff} = 
\TT_{p-1} \int \ccc\wedge e^{\fff} = S_{wz}\, ,
\ee
using eq.\refb{efer8}.

This shows that $S_{DBI}+S_{WZ}$ reduces to $S_{dbi}+S_{wz}$ under the 
identification \refb{efer21}. In principle we also need to check that 
any solution of the 
equations of motion derived from $S_{dbi}+S_{wz}$ is automatically a 
solution of the equations of motion derived from $S_{DBI}+S_{WZ}$. 
Presumably this can be done following the analysis of section \ref{s3}, 
but we have not worked out all the details.

Finally, we need to check that the supersymmetry transformations 
\refb{efer16} are compatible with the supersymmetry transformations 
\refb{efer6}. For this we need to calculate $\delta_{p-1} A_\mu$, 
$\delta_{p-1} Y^I$ and $\delta_{p-1} T$ using \refb{efer16}, \refb{efer21} 
and compare them with \refb{efer6}. The calculation is straightforward, 
and we get:
\ben \label{efer51}
&& \delta_p A_x = \delta_{p-1} A_x + \bar\epsilon \Gamma_{11} \Gamma_p 
\theta, \qquad \delta_p A_\alpha = \delta_{p-1} A_\alpha - \bar\epsilon 
\Gamma_{11} \Gamma_p
\theta \p_\alpha t, \qquad 
\nonumber \\ &&
\delta_p 
Y^I = \delta_{p-1} Y^I, \qquad 
\delta_p T = \delta_{p-1} T\, .
\een
Thus we see that $\delta_p$ and $\delta_{p-1}$ differ for the 
transformation laws of $A_x$ and $A_\alpha$. This difference, however, is 
precisely of the form induced by the function $\phi(x,\xi)$ in 
eq.\refb{egen1} with $\phi(x,\xi) =\bar\eps \Gamma_{11} \Gamma_p 
\theta(\xi)$. As was argued below \refb{egen1}, this is a 
gauge transformation. Thus we see that the action of $\delta_p$ and 
$\delta_{p-1}$ differ by a gauge transformation in the world-volume 
theory on the D-$p$-brane.

This establishes that the world
volume action on the kink reduces to that on a D-$(p-1)$-brane. The latter
has a local 
$\kappa$-symmetry which can 
be used to gauge away half 
of the world-volume fermion 
fields\cite{9610148,9610249,9611159,9611173,9612080}. 
This leads to a puzzle. Whereas 
on a
BPS D-brane the local $\kappa$-symmetry is postulated to be a gauge
symmetry, {\it i.e.} different configurations related by
$\kappa$-transformation are identified, on a kink solution
the appearance of the $\kappa$-symmetry seems accidental and {\it a
priori} there is no reason to identify field configurations which are
related by $\kappa$-symmetry. We believe the resolution of this puzzle
lies in the general principle advocated below \refb{egen1}
that any local transformation of the fields which does not change the 
action must be a gauge symmetry. This will
automatically imply that the $\kappa$-transformation is a gauge
transformation and we should identify the configurations related by
$\kappa$-transformation. This $\kappa$-symmetry can now be used to gauge 
away half of the fermion fields on the world-volume of the kink.

\sectiono{Vortex Solution on the Brane-Antibrane Pair} \label{s4}

In this section we shall generalize the construction of section \ref{s2} 
to a vortex solution on a brane-antibrane pair. For this we need to begin 
with a tachyon effective action on a brane-antibrane pair. In this case 
we have a complex tachyon field $T$, besides the massless gauge 
fields $A^{(1)}_\mu$, $A^{(2)}_\mu$ and scalar fields $Y_{(1)}^I$, 
$Y_{(2)}^I$ corresponding to the transverse coordinates of individual 
branes. We shall work with the following effective action that generalizes 
\refb{ez1}:\footnote{As in section \ref{s2}, we expect our analysis to be 
valid for a more general action of the form:
$$ -\int d^{p+1} x \, V(T, Y^I_{(1)}-Y^I_{(2)}) \, 
\Big[ \sqrt{-\det(g^{(1)}_{\mu\nu}+F^{(1)}_{\mu\nu})} 
F(G_{(1)}^{\mu\nu} D_\mu T^* D_\nu T) + 
\sqrt{-\det(g^{(2)}_{\mu\nu}+F^{(2)}_{\mu\nu})}
F(G_{(2)}^{\mu\nu} D_\mu T^* D_\nu T) \Big] $$
where $g^{(i)}_{\mu\nu}= \eta_{\mu\nu} + \p_\mu Y^I_{(i)} \p_\nu 
Y^I_{(i)}$ is the induced closed string metric on the $i$th 
brane, $G_{(i)}^{\mu\nu}$ is the open string metric on the $i$th 
brane and the function $F(u)$ grows as $u^{1/2}$ 
for large $u$.}$^,$\footnote{There have been various other proposals for 
the tachyon effective action and / or vortex solutions on brane-antibrane 
pair, see {\it e.g.} 
\cite{0012198,0012210,0211180,0007078,0008023}.} 
\be \label{evor1}
S = -\int d^{p+1} x \, V(T, Y^I_{(1)}-Y^I_{(2)}) \, \Big(\sqrt{-\det 
\bA_{(1)}} 
+ \sqrt{-\det 
\bA_{(2)}}\Big) \, ,
\ee
where
\be \label{evor2}
\bA_{(i)\mu\nu} = \eta_{\mu\nu} + F^{(i)}_{\mu\nu} + \p_\mu Y^I_{(i)} \p_\nu 
Y^I_{(i)} + {1\over 2} (D_\mu T)^* (D_\nu T) + {1\over 2} (D_\nu T)^* 
(D_\mu T)\, ,
\ee
\be \label{evor3}
F^{(i)}_{\mu\nu} = \p_\mu A^{(i)}_\nu - \p_\nu A^{(i)}_\mu\, ,
\qquad D_\mu T = (\p_\mu - i A^{(1)}_\mu + i A^{(2)}_\mu) T\, ,
\ee
and the potential $V(T)$ depends on $|T|$ and $\sum_I 
(Y^I_{(1)}-Y^I_{(2)})^2$ only. For small $T$, $V$ behaves as
\be \label{evor3a}
V(T,Y^I_{(1)}-Y^I_{(2)}) = \TT_p\, \left[ 1 + {1\over 2}\,\left\{ \sum_I 
\left({Y^I_{(1)}-Y^I_{(2)}\over2\pi}\right)^2 - 
{1\over 2}\right\}\, 
|T|^2 + \OO(|T|^4)\right]\, .
\ee 
$\TT_p$ denotes the tension of the individual D-$p$-branes.
Although this 
action 
has not been derived from first principles, we note that this obeys the 
following consistency conditions:
\begin{enumerate}

\item The action has the required invariance under the gauge 
transformation:
\be \label{evor4}
T \to e^{2i\alpha(x)}\, T, \qquad A^{(1)}_\mu \to A^{(1)}_\mu + \p_\mu 
\alpha(x), \qquad A^{(2)}_\mu \to A^{(2)}_\mu - \p_\mu
\alpha(x)\, .
\ee

\item For $T=0$ the action reduces to the sum of the usual 
DBI action on the individual branes.

\item If we require the fields to be invaiant under the symmetry 
$(-1)^{F_L}$ that exchanges the brane and the antibrane, we get the 
restriction:
\be \label{evor5}
T = \hbox{real}, \qquad A^{(1)}_\mu = A^{(2)}_\mu\equiv A_\mu \, , \qquad 
Y^I_{(1)} = 
Y^I_{(2)}\equiv Y^I\, .
\ee
Under this restriction the action becomes proportionl to that on a 
non-BPS D-$p$-brane, as given in \refb{ez1}. This is a necessary 
consistency check, as modding out a brane-antibrane configuration by 
$(-1)^{F_L}$ is supposed to produce a non-BPS D-$p$-brane\cite{9904207}.
\end{enumerate}
We should keep in mind however that these constraints do not fix the 
form of the action uniquely. Nevertheless we shall make the specific 
choice given in \refb{evor1} and proceed to study the vortex solution in 
this 
theory.

The energy momentum tensor $T^{\mu\nu}$ associated with this action is 
given by:
\be \label{evor5a}
T^{\mu\nu} = - V(T,Y^I_{(1)}-Y^I_{(2)}) \Big[\sqrt{-\det(\bA_{(1)})} 
(\bA_{(1)}^{-1})^{\mu\nu}_S
+ \sqrt{-\det(\bA_{(2)})} (\bA_{(2)}^{-1})^{\mu\nu}_S \Big]\, .
\ee
In order to construct a vortex solution we begin with the 
ansatz:
\be \label{evor6}
T(\rrho,\theta) = \bar f(\rrho) e^{i\theta}, \qquad A^{(1)}_\theta = 
-A^{(2)}_\theta = {1\over 2} \bar g(\rrho), 
\ee
where $\rrho$ and $\theta$ denote the polar coordinates in the $(x^{p-1}, 
x^p)$ plane, and $\bar f(\rrho)$ and $\bar g(\rrho)$ are real functions of 
$\rrho$ 
satisfying the boundary conditions:
\be \label{evor7}
\bar f(0)=0, \quad 
\bar f(\infty)=\infty, \quad \bar g(0)=0, \quad \bar g'(0) = 0\, .
\ee
All other fields vanish.
For such a background:
\be \label{evor8}
D_\rrho T = \bar f'(\rrho) e^{i\theta}, \qquad D_\theta T = i \bar f(\rrho) 
(1 - \bar g(\rrho)) e^{i\theta}, \qquad F^{(1)}_{\rrho\theta} = - 
F^{(2)}_{\rrho\theta} = {1\over 2} \bar g'(\rrho)\, .
\ee
Also in the polar coordinate that we have been using:
\be \label{evor9}
\eta = \pmatrix{\eta_{\alpha\beta} & & \cr
& 1 & \cr & & \rrho^2}\, , \qquad 0\le\alpha, \beta \le (p-2)\, .
\ee
This gives,
\be \label{evortex9a}
\bA_{(1)} = \pmatrix{\eta_{\alpha\beta} && \cr & 1 + (\bar f')^2 & {1\over 
2} \bar g' \cr & -{1\over 2} \bar g' & \rrho^2 + \bar f^2 (1 - \bar 
g)^2}\, , 
\qquad
\bA_{(2)} = \pmatrix{\eta_{\alpha\beta} && \cr & 1 + (\bar f')^2 & -{1\over
2} \bar g' \cr & {1\over 2} \bar g' & \rrho^2 + \bar f^2 (1 - \bar
g)^2}\, .
\ee
\be \label{evortex9b}
-\det(\bA_{(1)}) = -\det(\bA_{(2)}) = \Big[ \{1 + (\bar
f')^2\} \{
\rrho^2 + \bar f^2 (1 - \bar
g)^2 \} +
{1\over 4} (\bar g')^2 \Big]\, ,
\ee 
\ben \label{evor10}
T_{\alpha\beta} &=& -2\eta_{\alpha\beta} \, V(T) \, \sqrt{ \{1 + (\bar 
f')^2\} \{ 
\rrho^2 + \bar f^2 (1 - \bar 
g)^2 \} + 
{1\over 4} (\bar g')^2} \, , \nonumber \\
T_{\rrho\rrho} &=& - 2 V(T) \, \{\rrho^2 + \bar f^2 (1-\bar g)^2\} \Big/ 
\sqrt{\{1 + 
(\bar f')^2\} \{
\rrho^2 + \bar f^2 (1 - \bar
g)^2 \} +
{1\over 4} (\bar g')^2} \, , \nonumber \\
T_{\theta\theta} &=& - 2 V(T)\, \{1 + (\bar f')^2\} \Big/ \sqrt{\{1 + 
(\bar 
f')^2\} 
\{
\rrho^2 + \bar f^2 (1 - \bar
g)^2 \} +
{1\over 4} (\bar g')^2} \, ,
\een
where we have used the shorthand notation $V(T)$ to denote $V(T,0)$.
All other components of $T_{\mu\nu}$ vanish. The energy momentum 
conservation
\be \label{evor11}
0 = \p^\mu T_{\mu\rrho} = \p_\rrho T_{\rrho\rrho}\, ,
\ee
now shows that $T_{\rrho\rrho}$ must be a constant. 
Since $V(T)=V(\bar f e^{i\theta})$ falls off exponentially for large 
$|T|$, we see from
\refb{evor10} that $T_{\rrho\rrho}$ vanishes at $\infty$, 
unless $\bar g(\rrho)$ blows up sufficiently fast.
Shortly, we shall see that
$\bar g$ varies monotonically between 0 and 1, and hence is bounded.
This leads to the conclusion that $T_{\rrho\rrho}$ does vanish at infinity, 
and hence 
must be 
zero everywhere due to the conservation law \refb{evor11}.

To see that $\bar g(\rrho)$ varies monotonically between 0 and 1, we 
proceed as follows.
As a consequence of the equations of motion of 
the gauge fields, the 
$(p-2)$-dimensional  
energy density $\int \, \rrho\, d\rrho d\theta\, T_{00}$, with $T_{00}$ 
given 
in \refb{evor10}, must be minimized
with respect to the function $\bar g(\rrho)$ subject to the boundary 
condition 
\refb{evor7}.
Now if $\bar g(\rrho)$ exceeds 1 for 
some range of $\rrho$, then we can lower $T_{00}$ in that range by 
replacing the original $\bar g(\rrho)$ by another continuous 
function which is equal to the original function when the latter is less 
than 1, and which is 
equal to 1 when the latter exceeds 1. Thus the original $\bar g(\rrho)$ 
does not minimize energy and hence is not a solution of the equations 
of motion. This shows that a solution 
of the equations of motion must have $\bar g(\rrho)\le 1$ everywhere.
An exactly similar argument can be used to show that $\bar g(\rrho)\ge 0$ 
everywhere.
Furthermore, if $\bar g(\rrho)$ is not a monotone increasing function, 
then it will have a local maximum at some point $a$. We can now define 
a range $(a,b)$ on the $\rrho$ axis such that $\bar g(\rrho)<\bar 
g(a)$ for $a<\rrho<b$. ($b$ could be infinity.) 
In this case we can lower the energy of the configuration by replacing the 
original function by 
another continuous function that agrees with the original function outside 
the range $(a,b)$ and is equal to $\bar g(a)$ in the range 
$(a,b)$. Since this should not be possible if the original $\bar g(\rrho)$ 
is a 
solution of the equations of motion, we see that a solution of the 
equations of motion must have a 
monotone increasing $\bar g(\rrho)$.

Vanishing of $T_{\rrho\rrho}$ requires that
for every value of $\rrho$, either the numerator 
in the expression for $T_{\rrho\rrho}$
vanishes, which requires $V(T)$ to vanish, or the denominator 
blows 
up, 
which requires $\bar f'$ and/or $\bar g'$ to be infinite. 
$V(T)$ is finite at $\rrho=0$ where $T$ vanishes, thus it is not zero 
everywhere. 
Thus at least for $\rrho=0$, $\bar f'$ and/or $\bar g'$ must be infinite.
In 
analogy with the kink solution, we look for $\bar f$ and $\bar g$ of the 
form:
\be \label{evor12}
\bar f(\rrho) = f(a\rrho), \qquad \bar g(\rrho) = g(a\rrho)\, ,
\ee
and at the end take $a\to \infty$ limit, keeping the functions $f$ and $g$ 
fixed. The boundary conditions \refb{evor7} now translate to
\be \label{evor13}
f(0) = 0, \qquad f(\infty) = \infty, \qquad g(0)=0, \qquad 
g'(0) = 0\, .
\ee
We shall also impose the condition
\be \label{efpu}
f'(u) > 0 \qquad \hbox{for} \quad 0\le u < \infty\, .
\ee
This guarantees that $\bar f'(\rrho)=af'(a\rrho)$ is infinite
everywhere in the $a\to \infty$ limit. Once we have chosen $\bar f$ this 
way, 
we do not need to take $\bar g$ in 
the form given in \refb{evor12}. But this form allows for more general 
possibilities since 
without this the term involving $\bar g'$ will simply drop out in the 
scaling 
limit $a\to\infty$. On the other hand, by allowing $\bar g$ to scale as in 
\refb{evor12}
we do not 
preclude the case where $\bar g$ approaches a finite function in the 
$a\to\infty$ limit, since this will just correspond to choosing 
$g(\rrho)\equiv \bar g (\rrho/a)$ 
to be a nearly constant function except for very large $\rrho$.

Substituting \refb{evor12} into \refb{evortex9b}, \refb{evor10} we get,
for large $a$,
\be \label{evordet}
-\det(\bA_{(1)}) = -\det(\bA_{(2)}) \simeq a^2 \, (f'(a\rrho))^2 \, \left[
\rrho^2 + f(a\rrho)^2 (1 - g(a\rrho))^2 +{1\over 4} 
(g'(a\rrho)/f'(a\rrho))^2\right]\, ,
\ee
and,
\be \label{evor14}
T_{\alpha\beta} \simeq -2\eta_{\alpha\beta} \, V(f(a\rrho)) \, a \, 
f'(a\rrho) \, 
\sqrt{\rrho^2 + f(a\rrho)^2 (1 - g(a\rrho))^2 +{1\over 4} (g'(a\rrho)
/f'(a\rrho))^2}\, ,
\ee
\be \label{evor14a}
T_{\rrho\rrho} \simeq - 2 V(f(a\rrho)) {\rrho^2 +f(a\rrho)^2 (1 - 
g(a\rrho))^2\over a 
f'(a\rrho) \sqrt{\rrho^2 + f(a\rrho)^2 (1 - g(a\rrho))^2 +{1\over 4} 
(g'(a\rrho)
/f'(a\rrho))^2}} \, .
\ee
Thus $T_{\rrho\rrho}$ vanishes everywhere in the $a\to\infty$ limit as 
required. On the other hand,   
integrating \refb{evor14} over the $(\rrho,\theta)$ coordinates gives the 
$(p-2+1)$ 
dimensional energy momentum tensor 
$T^{vortex}_{\alpha\beta}$ on the vortex:
\be \label{evor15}
T^{vortex}_{\alpha\beta} = -4\pi\eta_{\alpha\beta} \, \int_0^\infty d\rrho 
\, 
V(f(a\rrho)) \, a \, f'(a\rrho) \,
\sqrt{\rrho^2 + f(a\rrho)^2 (1 - g(a\rrho))^2 + {1\over 4} (g'(a\rrho)
/f'(a\rrho))^2}\, .
\ee
Defining:
\be \label{evor16}
y = f(a\rrho), \qquad \wh\rrho(y) = a^{-1} 
f^{-1}(y)\, , \qquad \wh g(y) = g(a\rrho) = g(a\wh\rrho(y)), 
\ee
where $f^{-1}$ denotes the inverse function of $f$, 
we can rewrite \refb{evor15} as
\be \label{evor17}
T^{vortex}_{\alpha\beta} = -4\pi\eta_{\alpha\beta} \, \int_0^\infty \, dy 
\, 
V(y)\,  
\sqrt{\wh\rrho(y)^2 + y^2 \{1 - \wh g(y)\}^2 + {1\over 4} \wh 
g'(y)^2}\, .
\ee
{}From \refb{evor16} it follows that in the $a\to\infty$ limit, 
$\wh\rrho(y)$ vanishes for any finite $y$. Thus \refb{evor17} further 
simplifies to:
\be \label{evor18}
T^{vortex}_{\alpha\beta} = -4\pi\eta_{\alpha\beta} \, \int_0^\infty \, dy 
\,
V(y)\,
\sqrt{y^2 \{1 - \wh g(y)\}^2 + {1\over 4} \wh g'(y)^2}\, .
\ee
We now see that as in the case of the kink solution, \refb{evor18} is 
completely insensitive to the choice of the function $f$, although it does 
depend on the choice of $\wh g(y)$. $\wh g(y)$ in turn is determined by 
the 
equations of motion of the gauge fields, or equivalently, by minimizing 
the 
expression for the energy $T^{vortex}_{00}$, subject to the boundary 
conditions:
\be \label{evor19}
\wh g(0)=0, \qquad \wh g'(0) = 0\, .
\ee
This leads to the following differential equation for $\wh g(y)$:
\ben \label{evor19a}
&& {1\over 4} \p_y \left[ V(y) \, \wh g'(y) \Big/ \sqrt{y^2 \{1 - \wh 
g(y)\}^2 + 
{1\over 4} \wh 
g'(y)^2}\right] \nonumber \\
&&
+ V(y) \, y^2 (1-\wh g(y)) \Big/ \sqrt{y^2 \{1 - \wh g(y)\}^2 + {1\over 
4} \wh 
g'(y)^2} = 
0\, . 
\een
Thus
$\wh g(y)$ and 
the final expression for 
$T^{vortex}_{\alpha\beta}$ 
are
determined completely in terms of the potential $V(T)$, independently of 
the choice of the function $f$.\footnote{The choice $f(a\rrho)=a\rrho$ gives 
$T=a(x^{p-1}+i x^p)$ in the cartesian coordinate system. This resembles 
the vortex solution in boundary string field 
theory\cite{0012198,0012210}. However, unlike in \cite{0012198,0012210}, 
here we have background gauge fields present. This is not necessarily a 
contradiction, since the fields used here could be related to those 
in \cite{0012198,0012210} by a non-trivial field redefinition.
In fact, we would like to note that generically, when both the real and 
the imaginary parts of the tachyon are non-zero and are not proportional 
to each other, we have a source for the gauge field 
$A_\mu^{(1)}-A_\mu^{(2)}$, and hence it is not possible to find a solution 
of the equations of motion keeping the gauge fields to zero. Boundary 
string field theory seems to use a very special definition of fields where 
this is possible in the $a\to\infty$ limit.} Furthermore, as in the case 
of the kink 
solution, most of the contribution to $T^{vortex}_{\alpha\beta}$ comes 
from a 
finite range of values of $y$, which corresponds to a region in $\rrho$ 
space of width $1/a$ around the origin. Thus in the $a\to 0$ limit, 
$T_{\alpha\beta}$ has the form of a $\delta$-function centered around the 
origin of the $(x^{p-1}, x^p)$ plane:
\be \label{evor20}
T_{\alpha\beta} = -4\pi\eta_{\alpha\beta} \, \delta(x^{p-1}) \, 
\delta(x^p) 
\, \int_0^\infty \, dy \,
V(y)\,
\sqrt{ y^2 \{1 - \wh g(y)\}^2 + {1\over 4} \wh g'(y)^2}\, .
\ee
This agrees with the identification of the vortex solution as a 
D-$(p-2)$-brane, as for the latter the energy-momentum tensor is localised 
on a $(p-2)$-dimensional surface. (This can be seen by examining 
the boundary state describing a D-$(p-2)$-brane.) The tension of the 
D-$(p-2)$-brane is identified as:
\be \label{evor21}
\TT_{p-2} = 4\pi\, \int_0^\infty \, dy \,
V(y)\,
\sqrt{ y^2 \{1 - \wh g(y)\}^2 + {1\over 4} \wh g'(y)^2}\, .
\ee

Before concluding this section, we shall determine the asymptotic 
behaviour of $\wh g(y)$ satisfying eqs.\refb{evor19} and 
\refb{evor19a}.
Our previous arguments for the function $\bar g(\rrho)$, when translated 
to $\wh g(y)$, shows that $\wh g(y)$ must be a monotone increasing 
function of $y$, and must lie between 0 and 1. The boundary condition 
forces $\wh g(y)$ to vanish at $y=0$. We shall now show that given a 
mild constraint on the potential $V(T)$,
$\wh g(y)$ must approach 1 as $y\to\infty$. We shall 
begin by assuming 
that $\wh g(y)$ approaches some constant value $(1-C)$ as 
$y\to\infty$, and then show that $C$ must vanish. If $C\ne 0$, then the 
dominant term inside the square root for large $y$ is the first term 
which takes the value $y^2 C^2$, since $\wh g'(y)$ vanishes for large $y$. 
Thus for large $y$, \refb{evor19a} takes 
the form:
\be \label{easym1}
{1\over 4} \p_y \left[ V(y) \, \wh g'(y) / yC\right] + y V(y) = 0\, .
\ee
Since $\p_y\big(\wh g'(y) / yC\big)$ approaches 0 as $y\to\infty$, clearly 
the only part 
of the first term in \refb{easym1} that can possibly cancel the second 
term is 
$V'(y) \wh g'(y) / (4 y C)$. If this has to cancel the second term, we 
require
\be \label{easym2}
V'(y) / V(y) \simeq - 4 y^2 C / \bar g'(y)\, , \qquad \hbox{for large 
$y$.}
\ee
Since $\wh g(y)$ approaches a constant as $y\to\infty$, $\wh g'(y)$ must 
fall off faster than $1/y$ for large $y$. Thus the magnitude 
of the right hand side of \refb{easym2}
increases faster than $y^3$ for large $y$. This, in turn, shows that 
$-V'(y) / V(y)$ 
must also increase faster than $y^3$ for large $y$.
Neither a potential of the form $e^{-\beta y}$ obtained from the analysis 
of time dependent solutions\cite{0204143}, nor a potential of the form 
$e^{-\beta y^2}$ given by boundary string field 
theory\cite{0012198,0012210} satisfies this condition. Thus our original 
assumption must be wrong and $C$ must vanish for either of these choices 
of $V(T)$.

This leads us to the conclusion that if $-V'(y)/V(y)$ does not increase 
faster 
than $y^3$ for large $y$, we must have
\be \label{easym3}
\lim_{y\to\infty} \wh g(y) = 1\, .
\ee
This, in turn, has the following consequence. From \refb{evor8}, 
\refb{evor16}, \refb{easym3} we have
\be \label{easym4}
\int d\rrho \, d\theta \, (F^{(1)}_{\rrho\theta} - F^{(2)}_{\rrho\theta}) = 
2\pi (\bar g(\infty) - \bar g(0)) = 2\pi (\wh g(\infty) - \wh g(0)) = 
2\pi\, .
\ee
This answer is universal, independent of the choice of the potential 
$V(T)$, 
provided $V(T)$ satisfies the mild asymptotic condition given above 
\refb{easym3}. This is also the same answer that we would have gotten if 
we had 
a usual abelien Higgs model with an action given by the sum of a kinetic 
and a potential term. Finally, for this gauge field background, if we 
compute the Ramond-Ramond (RR) charge of the vortex using the usual 
coupling between the 
world-volume gauge fields and the RR fields at zero tachyon background, we 
get the correct expression for the RR charge of the vortex. Thus the net
additional contribution to the RR charge from the tachyon dependent 
coupling of the RR 
fields\cite{9905195} must vanish. This is in contrast with the boundary 
string field 
theory result\cite{0012198,0012210} where the complete contribution to the 
RR charge 
comes from the tachyon fields. This again reflects that the fields used 
here are related to those in boundary string field theory by non-trivial 
field redefinition.

\sectiono{World-volume Action on the Vortex} \label{s5}

We shall now study the world-volume action on the vortex. We begin by
introducing some notation. We shall denote by $x^i$ for $(p-1)\le i\le p$
the coordinates transverse to the world-volume of the vortex but
tangential to the original brane and by $\xi^\alpha$ for $0\le\alpha\le
(p-2)$ the coordinates tangential to the vortex. We shall express the
classical vortex solution of \refb{evor6} in cartesian coordinates as
\be \label{ew1}
A^{(1)}_i = - A^{(2)}_i = \bar h_i(\vec x), \qquad T(\vec x) = \bar 
f(\vec 
x)\, ,
\ee
where 
\be \label{ew2}
\bar h_{p-1}(\vec x) = -{x^p\over 2\rrho^2} \bar g(\rrho), \quad \bar 
h_p(\vec 
x) = {x^{p-1}\over 2\rrho^2} \bar g(\rrho), \quad \bar f(\vec x) = \bar 
f(\rrho), \quad \rrho\equiv 
|\vec x|\, , \quad \vec x=(x^{p-1}, x^p)\, .
\ee
We now make the following ansatz for the fluctuating fields on the 
world-volume of the vortex:
\ben \label{ew3}
&& A^{(1)}_i(\vec x, \xi) = \bar h_i(\vec x - \vec t(\xi))\, , \qquad 
A^{(2)}_i(\vec x, \xi) = -\bar h_i(\vec x -\vec t(\xi))\, , \nonumber \\
&& A^{(1)}_\alpha(\vec x, \xi) =-\bar h_i(\vec x -\vec t(\xi)) \p_\alpha 
t^i + a_\alpha(\xi), \qquad A^{(2)}_\alpha(\vec x, \xi) =\bar h_i(\vec x 
-
\vec t(\xi)) \p_\alpha
t^i + a_\alpha(\xi), \nonumber \\
&& Y^I_{(1)}(\vec x, \xi) = Y^I_{(2)}(\vec x, \xi) = y^I(\xi), \qquad 
T(\vec x, \xi) = \bar f(\vec x - \vec t(\xi))\, .
\een
Thus the world volume fields on the vortex are $y^I(\xi)$, $t^i(\xi)$ and 
$a_\alpha(\xi)$. 

We shall now substitute this ansatz into the action \refb{evor1} 
and evaluate the action.
Using the ansatz \refb{ew3} and the definitions \refb{evor3} we get
\ben \label{ew4}
&& D_i T = \p_i \bar f - 2 i \bar h_i \bar f \equiv D_i\bar f\, , \qquad
D_\alpha T = -D_i \bar f \p_\alpha t^i \, , \nonumber \\ 
&& F^{(1)}_{ij} = (\p_i \bar h_j - \p_j \bar h_i) \, , \qquad 
F^{(2)}_{ij} = -(\p_i \bar h_j - \p_j \bar h_i) \, , \nonumber \\
&& F^{(1)}_{i\beta} = - F^{(2)}_{i\beta} = -(\p_i \bar h_j - \p_j \bar 
h_i) \p_\beta t^j\, , \qquad 
F^{(1)}_{\alpha j} = - F^{(2)}_{\alpha j} = -(\p_i \bar h_j - \p_j \bar
h_i) \p_\alpha t^i\, , \nonumber \\
&& F^{(1)}_{\alpha \beta} = f_{\alpha \beta} + (\p_i \bar h_j - \p_j \bar
h_i) \p_\alpha t^i \p_\beta t^j\, , \qquad
F^{(2)}_{\alpha \beta} = f_{\alpha \beta} - (\p_i \bar h_j - \p_j \bar
h_i) \p_\alpha t^i \p_\beta t^j\, , \nonumber \\
\een
where
\be \label{ew5}
f_{\alpha\beta} = \p_\alpha a_\beta - \p_\beta a_\alpha\, .
\ee
In each expression the arguments of $\bar h_i$ and $\bar f$ are $(\vec 
x-\vec t(\xi))$ which we have suppressed.
{}From \refb{evor2} we now get
\ben \label{ew6}
&& \bA_{(1)ij} = \delta_{ij} + (\p_i \bar h_j - \p_j \bar
h_i) +{1\over 2} \Big((D_i\bar f)^* D_j\bar f + (D_j \bar f)^* D_i\bar 
f\Big) \nonumber \\
&& \bA_{(1)i\beta} = -(\p_i \bar h_j - \p_j \bar
h_i) \p_\beta t^j - {1\over 2} \Big((D_i\bar f)^* D_j\bar f + (D_j \bar 
f)^* D_i\bar
f\Big) \p_\beta t^j \nonumber \\
&& \bA_{(1)\alpha j}  = -(\p_i \bar h_j - \p_j \bar
h_i) \p_\alpha t^i - {1\over 2} \Big((D_i\bar f)^* D_j\bar f + (D_j \bar
f)^* D_i\bar
f\Big) \p_\alpha t^i\nonumber \\
&& \bA_{(1)\alpha \beta} = \eta_{\alpha\beta} + f_{\alpha\beta} 
+ \p_\alpha y^I \p_\beta y^I + (\p_i 
\bar h_j - \p_j \bar
h_i) \p_\alpha t^i \p_\beta t^j \nonumber \\
&& \qquad \qquad + {1\over 2} 
\Big((D_i\bar f)^* D_j\bar f 
+ (D_j \bar
f)^* D_i\bar
f\Big) \p_\alpha t^i \p_\beta t^j  \nonumber \\
&& \bA_{(2)ij} = \delta_{ij} - (\p_i \bar h_j - \p_j \bar
h_i) +{1\over 2} \Big((D_i\bar f)^* D_j\bar f + (D_j \bar f)^* D_i\bar 
f\Big) \nonumber \\
&& \bA_{(2)i\beta} = (\p_i \bar h_j - \p_j \bar
h_i) \p_\beta t^j - {1\over 2} \Big((D_i\bar f)^* D_j\bar f + (D_j \bar 
f)^* D_i\bar
f\Big) \p_\beta t^j \nonumber \\
&& \bA_{(2)\alpha j}  = (\p_i \bar h_j - \p_j \bar
h_i) \p_\alpha t^i - {1\over 2} \Big((D_i\bar f)^* D_j\bar f + (D_j \bar
f)^* D_i\bar
f\Big) \p_\alpha t^i\nonumber \\
&& \bA_{(2)\alpha \beta} = \eta_{\alpha\beta} + f_{\alpha\beta} + 
\p_\alpha y^I \p_\beta y^I - (\p_i 
\bar h_j - \p_j \bar
h_i) \p_\alpha t^i \p_\beta t^j \nonumber \\
&& \qquad \qquad + {1\over 2} \Big((D_i\bar 
f)^* D_j\bar f 
+ (D_j \bar
f)^* D_i\bar
f\Big) \p_\alpha t^i \p_\beta t^j \, . \nonumber \\
\een
We now simplify the computation of the determinants by subtracting 
appropriate multiples of the first two rows/columns from the rest of the 
rows / columns. This does not change the determinant of the matrix. More 
precisely, we define:
\ben \label{ew7}
&& \wh \bA_{(s)\alpha\nu} = \bA_{(s)\alpha\nu} 
+ \bA_{(s)i\nu}\p_\alpha 
t^i, \qquad \wh \bA_{(s)i\nu} = \bA_{(s)i\nu}\, , \nonumber \\
&& \wt \bA_{(s)\mu\beta} = \wh \bA_{(s)\mu\beta} + \wh 
\bA_{(s)\mu j} \p_\beta t^j, \qquad
\wt \bA_{(s)\mu j} = \wh
\bA_{(s)\mu j}\, ,
\qquad \hbox{for} \quad 0\le\mu,\nu\le p\, .
\een
Under this transformation we have:
\be \label{ew8}
\det(\bA_{(s)}) = \det(\wh \bA_{(s)}) = \det(\wt \bA_{(s)}) \, , \qquad 
s=1,2
\, .
\ee
On the other hand, we have, from \refb{ew6}, \refb{ew7}
\ben \label{ew9}
&& \wt \bA_{(1)ij} = \delta_{ij} + (\p_i \bar h_j - \p_j \bar
h_i) +{1\over 2} \Big((D_i\bar f)^* D_j\bar f + (D_j \bar f)^* D_i\bar
f\Big) \nonumber \\
&& \wt \bA_{(1)i\beta} = \p_\beta t^i, \qquad \wt \bA_{(1)\alpha 
j}  = 
\p_\alpha t^j\, , \nonumber \\  
&& \wt \bA_{(1)\alpha \beta} = \eta_{\alpha\beta} + f_{\alpha\beta} + 
\p_\alpha y^I \p_\beta y^I + \p_\alpha t^i \p_\beta t^i \, , \nonumber \\
&& \wt \bA_{(2)ij} = \delta_{ij} - (\p_i \bar h_j - \p_j \bar
h_i) +{1\over 2} \Big((D_i\bar f)^* D_j\bar f + (D_j \bar f)^* D_i\bar
f\Big) \nonumber \\
&& \wt \bA_{(2)i\beta} = \p_\beta t^i, \qquad \wt \bA_{(2)\alpha j}  
=\p_\alpha t^j\, , \nonumber \\
&& \wt \bA_{(2)\alpha \beta} = \eta_{\alpha\beta} + f_{\alpha\beta} +
\p_\alpha y^I \p_\beta y^I + \p_\alpha t^i \p_\beta t^i \, , \nonumber \\
\een
Examining the form of the $ij$ component of the matrices $\wt \bA_{(1)}$ 
and 
$\wt \bA_{(2)}$ we see that they are precisely of the same form as one 
would 
get 
for the classical vortex solution without fluctuation, except for the 
replacement of $\vec x$ 
by 
$(\vec x - \vec t(\xi))$ in the argument of $\bar h_i$ and $\bar f$. Since 
this determinant given in \refb{evordet} has an explicit factor of 
$a^2$
which becomes large in the $a\to\infty$ limit, 
and since $\wt\bA_{(s)i\beta}$, $\wt\bA_{(s)\alpha j}$ and 
$\wt\bA_{(s)\alpha\beta}$ are all of order one, 
in this limit we can ignore 
the contribution from the off-diagonal elements $\wt \bA_{(s)i\beta}$ 
and 
$\wt \bA_{(s)\alpha j}$ in evaluating $\det(\wt\bA_{(s)})$. Thus the 
resulting action is given by:
\ben \label{ew10}
&& - 2 \int d^{p-1} \xi \int d\rrho d\theta \, V(f(a\rrho)) \,
a \,
f'(a\rrho) \,
\sqrt{\rrho^2 + f(a\rrho)^2 (1 - g(a\rrho))^2 +{1\over 4} (g'(a\rrho)
/f'(a\rrho))^2} \nonumber \\
&& ~ \qquad \qquad \times
\,  \sqrt{-\det \AAA}\, ,
\een
where
\be \label{ew11}
\AAA_{\alpha\beta} = \eta_{\alpha\beta} + f_{\alpha\beta} +
\p_\alpha y^I \p_\beta y^I + \p_\alpha t^i \p_\beta t^i \, .
\ee
In \refb{ew10} we have redefined $\rrho$ to be $|\vec x -\vec t(\xi)|$, 
and 
$\theta$ to be $\tan^{-1} \Big({x^{p-1} - t^{p-1}(\xi) \over x^{p} - 
t^{p}(\xi)}\Big)$. We can now explicitly perform the $\rrho$ and $\theta$ 
integrals as in section \ref{s4} and use \refb{evor21} to rewrite the 
action \refb{ew10} as
\be \label{ew12}
- \TT_{p-2} \, \int d^{p-1} \xi \, \sqrt{-\det \AAA}\, .
\ee
This is precisely the world-volume action on a BPS D-$(p-2)$-brane with 
$t^i$ and $y^I$ interpreted as 
the coordinates transverse to the brane for $(p-1)\le i\le p$ and 
$(p+1)\le I\le 9$  and $a_\alpha$ interpreted as the 
gauge field on the D-brane world-volume.

As in section \ref{s3}, in order to establish completely that the 
dynamics of the world-volume theory on the vortex is governed by the 
action 
\refb{ew12} we need to show that given any solution of the equations of 
motion derived from this action, \refb{ew3} provides us with a solution of 
the full $(p+1)$-dimensional equations of motion. We have not checked 
this, but believe that this can be done following techniques similar to 
that discussed in section \ref{s3}.

\sectiono{Discussion} \label{s6}

In this paper we have analyzed kink and vortex solutions in tachyon
effective field theory by postulating suitable form of the tachyon
effective action on the non-BPS D-brane and brane-antibrane system
respectively. In both cases the topological soliton has all the right
properties for describing a BPS D-brane. These properties include
localization of the energy-momentum tensor on subspaces of codimensions 1
and 2 respectively, as is expected of a D-brane and also the DBI form of
the effective action describing the world-volume theory on the soliton. 
For the kink solution we have also done the analysis including the 
world-volume fermions, and shown the appearance of $\kappa$-symmetry in 
the world-volume theory on the kink.

One feature of both the solutions is infinite spatial gradient of the
tachyon field away from the core of the soliton. If we want to construct a
solution describing tachyon 
matter\cite{0203211,0203265,0204143,0202210,0209090,0301038,0302146,0208196}
in 
the
presence of such a soliton, then the spatial gradient of the tachyon field
represents local velocity of the tachyon matter\cite{0204143,0209122}. 
More 
precisely, the local $(p+1)$-velocity of the dust is given by $u_\mu = 
-\p_\mu T$. Thus large
positive gradient of the tachyon implies large local velocity {\it
towards} the core of the soliton. 
This shows that tachyon matter in the presence of such a solution will 
fall towards the core of the soliton.
If this feature survives in the full string theory, then it will imply 
that
any 
tachyon matter in contact with the soliton 
will be sucked in immediately.
This is consistent with 
the
analysis of \cite{0207105,0212248} where similar effect was found by 
analyzing the
boundary state associated with the time dependent solutions.\footnote{We 
should keep in mind, however, that this result is exact only for bosonic 
string theory. For superstring theory the corresponding boundary conformal 
field theory is not solvable, and hence no exact result can be obtained.} 
This might provide a very effective means of absorbing tachyon matter from 
the surrounding by a defect brane, and drastically modify the results of 
refs.\cite{0207119,0207156} for the formation of 
topological 
defects
during the rolling of the tachyon. The appearance of infinite slope during 
the dynamical process of defect formation has 
already been observed in \cite{0301101}.
We should note however that a  
different type of solution where a codimension 1 soliton and 
tachyon 
matter 
coexist has been constructed in 
\cite{0301179}.

Another surprising feature of both the kink and the vortex solutions is 
that the world-volume 
theory on the soliton has exactly the DBI form without any higher 
derivative corrections. This means that all such corrections must come 
from higher derivative corrections to the original actions \refb{ez1} and 
\refb{evor1}. This may seem accidental, but may be significant for the 
following reasons. 
This result suggests that there is a close relation between the systematic 
derivative (of field strength) expansion of the world-volume action of the 
non-BPS D-$p$-brane (D-$p$-brane - $\bd$-$p$-brane pair) and that of the 
BPS soliton solution representing D-$(p-1)$ brane 
(D-$(p-2)$-brane). 
It will be interesting 
to explore this line of thought to see if one can establish a precise 
connection between the two. Since the derivative expansion on the 
world-volume of BPS D-branes is well understood, finding a connection of 
the type mentioned above will provide a better understanding of the 
derivative expansion of the world-volume action of a non-BPS D-brane / 
brane-antibrane system.

One question that we have not addressed in this paper is the analysis of 
the world-volume theories on (multiple) kink-antikink pairs and 
multivortex solutions. The construction of these solutions should be quite 
straightforward following {\it e.g.} the analysis of 
\cite{0102174,0208217,0211180}. In a finite region around the location of 
each 
soliton the solution will have the form discussed in sections \ref{s2} 
and \ref{s4}, and we need to ensure that before taking the $a\to\infty$ 
limit, the various fields match smoothly, keeping $|T|$ or order $a$ or 
larger in the intervening space. Analysis of the world-volume theory 
around such a background will clearly yield the sum of the world-volume 
actions on the individual solitons, since essentially the field 
configurations around individual solitons do not talk to each other in the 
$a\to\infty$ limit. The interesting question is whether we can see the 
excitations associated with the fundamental string stretched between the 
solitons. We believe these excitations must come from classical solutions 
(`solitons') describing fundamental string along the line of 
refs.\cite{0002223,0009061,0010240,9901159}. 
We can, for example, take the solutions in the DBI theory given in 
\cite{9708147,9709027,9709014,9711094} and lift them to solutions of the 
equations of motion 
derived from \refb{ez1} or \refb{evor1} using \refb{ezsol10}, 
\refb{ezsol18} or \refb{ew3}.
The (spontaneously broken) 
gauge 
symmetry that mixes the states of the open string living on 
individual D-branes
with  states of the open string stretched between different D-branes, 
exchanges perturbative states with `solitonic' states, and hence is 
analogous to the electric magnetic duality symmetry in gauge 
theories\cite{mon-ol,wit-ol,osborn,9402032,9408074,9407087,9408099}.

The general lesson that one could learn from the results of this paper is 
that for many purposes, it is useful to complement the supergravity 
action, 
describing low energy effective action of closed string theory, by 
coupling it to the 
tachyon effective action of the type described in this paper. In such a 
theory, BPS D-branes arise naturally as topological solitons rather than 
having 
to be added by hand, and we get the correct { low energy} effective 
action on these D-branes. Furthermore, we have seen earlier that this 
effective 
action is capable of describing certain time dependent solutions of open 
string theory\cite{0203211,0203265,0204143}, and solutions describing the 
fundamental string\cite{0002223,0009061,0010240}. Coupling the tachyon 
field to 
supergravity does not give rise to any new perturbative physical states, 
and hence does not violate any known result in string theory. Finally, as 
was argued in \cite{0209122}, coupling of the tachyon effective action to 
gravity may resolve some of the conceptual problems involving `time' in 
quantum gravity. 

\bigskip

{\bf Acknowledgement}: I would like to thank D.~Choudhury, 
J.~Cline, D.~Ghoshal, 
R.~Gopakumar, 
D.~Jatkar, J.~Minahan, S.~Panda, L.~Rastelli and B.~Zwiebach for useful 
discussions and comments on the manuscript.

\end{document}